\documentclass[12pt,preprint]{aastex}

\begin{document}


\title{An Empirical Calibration of the Completeness of the SDSS Quasar Survey}


\author{
Daniel E.\ Vanden Berk\altaffilmark{1,2},
Donald P.\ Schneider\altaffilmark{2},
Gordon T.\ Richards\altaffilmark{3},
Patrick B.\ Hall\altaffilmark{3},
Michael A.\ Strauss\altaffilmark{3},
Robert Brunner\altaffilmark{4},
Xiaohui Fan\altaffilmark{5},
Ivan K.\ Baldry\altaffilmark{6}
Donald G.\ York\altaffilmark{7,8},
James E.\ Gunn\altaffilmark{3},
Robert C.\ Nichol\altaffilmark{9,10}
Avery Meiksin\altaffilmark{11},
Jon Brinkmann\altaffilmark{12}
}

\altaffiltext{1}{Department of Physics and Astronomy, University of Pittsburgh, 3941 O'Hara Street, Pittsburgh, PA 15260; danvb@bruno.phyast.pitt.edu.}
\altaffiltext{2}{Department of Astronomy and Astrophysics, The Pennsylvania State University, 525 Davey Laboratory, University Park, PA 16802; dps@astro.psu.edu.}
\altaffiltext{3}{Princeton University Observatory, Peyton Hall, Princeton, NJ 08544; gtr@astro.princeton.edu, pathall@astro.princeton.edu, strauss@astro.princeton.edu, jeg@astro.princeton.edu.}
\altaffiltext{4}{Department of Astronomy, University of Illinois at Urbana-Champaign, 1002 W. Green Street, Urbana, IL 61801 USA; rb@astro.uiuc.edu.}
\altaffiltext{5}{Steward Observatory, University of Arizona, 933 North Cherry Avenue, Tucson, AZ 85721; fan@as.arizona.edu.}
\altaffiltext{6}{Department of Physics and Astronomy, Johns Hopkins University, Baltimore, MD 21218; baldry@pha.jhu.edu.}
\altaffiltext{7}{Department of Astronomy and Astrophysics, The University of Chicago, 5640 South Ellis Avenue, Chicago, IL 60637; don@oddjob.uchicago.edu.}
\altaffiltext{8}{Enrico Fermi Institute, The University of Chicago, 5640 South Ellis Avenue, Chicago, IL 60637.}
\altaffiltext{9}{Physics Department, Carnegie Mellon University, 5000 Forbes Avenue, Pittsburgh, PA 15213.}
\altaffiltext{10}{Institute of Cosmology and Gravitation, University of Portsmouth, Portsmouth PO1 2EG, UK; bob.nichol@port.ac.uk.}
\altaffiltext{11}{Institute for Astronomy, University of Edinburgh, Blackford Hill, Edinburgh EH9, UK; a.meiksin@roe.ac.uk.}
\altaffiltext{12}{Apache Point Observatory, P.O.\ Box 59, Sunspot, NM 88349; jb@apo.nmsu.edu.}


\begin{abstract}
Spectra of nearly 20000 point-like objects to a Galactic reddening
corrected magnitude of $i=19.1$ have been obtained to test the
completeness of the Sloan Digital Sky Survey (SDSS) quasar survey.
We focus on spatially unresolved quasars, which comprise $94\%$ of
all SDSS quasars to the main survey magnitude limit.  The objects were
selected from all regions of color space, sparsely sampled from within a
$278\,{\rm deg}^2$ area (effective area $233\,{\rm deg}^2$) of sky covered
by this study.  Only ten quasars were identified that were not targeted
as candidates by the SDSS quasar survey (including both color and radio
source selection).  The inferred density of unresolved quasars on the
sky that are missed by the SDSS algorithm is $0.44\,{\rm deg}^{-2}$,
compared to $8.28\,{\rm deg}^{-2}$ for the selected quasar density,
giving a completeness of $94.9^{+2.6}_{-3.8}\%$ to the limiting magnitude.
Omitting radio selection reduces the color-only selection completeness by
about $1\%$.  Of the ten newly identified quasars, three have detected
broad absorption line systems, six are significantly redder than other
quasars at the same redshift, and four have redshifts between 2.7 and
3.0 (the redshift range where the SDSS colors of quasars intersect the
stellar locus).  The fraction of quasars (and other unresolved sources)
missed due to image defects and blends, independent of the selection
algorithm, is $\approx 4\%$, but this number varies by a few percent
with magnitude.  Quasars with extended images comprise about $6\%$
of the SDSS sample, and the completeness of the selection algorithm
for extended quasars is approximately $81\%$, based on the SDSS galaxy
survey.  The combined end-to-end completeness for the SDSS quasar survey
is $\approx89\%$.  The total corrected density of quasars on the sky to
$i=19.1$ is estimated to be $10.2\,{\rm deg}^{-2}$.  The SDSS completeness
restricted to $z\gtrsim3$ quasars is expected to be considerably lower,
and is a strong function of redshift.  The determination of the global
completeness is required for the statistical properties of quasars to
$i=19.1$ derived from the SDSS dataset, such as the luminosity function
and correlation function, to be accurately determined.  \end{abstract}

\keywords{quasars: general --- surveys}


\section{Introduction}

The Sloan Digital Sky Survey \citep[SDSS,][]{york00} is producing
the largest dataset of spectroscopically confirmed quasars to date,
with the goal of obtaining $\approx100,000$ unique quasar spectra
by survey end.  It is necessary to understand the survey selection
function, including the completeness, in order to use the dataset for
statistical purposes.  For example, an accurate measure of the quasar
luminosity function is dependent on the knowledge of the quasar
distribution in the color space from which the quasars were selected.
Public releases of the quasar dataset \citep{schneider02,schneider03}
have so far not been accompanied by a detailed description of
the selection function, nor of an estimate of the completeness.
That is because the first two published quasar catalogs did not use
the exact final SDSS quasar target selection algorithm described by
\citet{richards02}.  The goal of the current study is to quantify
the global completeness of the finalized algorithm, which has been
used since the Fall of 2002, and will be used for the remainder of
the survey.

Given the high precision of the five-band SDSS imaging photometry,
the design of the filters to optimally distinguish between stars
and quasars, and inclusion of optical source matching with the FIRST
radio source catalog \citep{becker95}, the completeness of the SDSS quasar
survey has been expected to be very high.  Preliminary estimates of
the quasar completeness have come from comparison of radio selected
and optical color selected quasar samples in the same regions of
the sky \citep{ivezic02}.  Assuming that quasars with detectable
radio emission are distributed in optical color space the same way
as is the quasar population in general, the radio comparison shows
that the completeness is likely to exceed $89\%$.  This result is
consistent with the estimates given by \citet{richards02}, who used
comparisons to previously identified quasars and simulated spectra
to conclude that the completeness should exceed $90\%$.  The quasars
missed by the SDSS algorithm were lost due mainly to image defects
and unresolved blends with other objects.  These analyses are not
definitive, however, because the region of color space containing
the vast majority of stars --- the stellar locus --- has not been
systematically searched for quasars.

In order to provide an empirical calibration of the completeness of the
SDSS spectroscopic targeting algorithms, we have undertaken a
spectroscopic survey of point sources to a dereddened $i$ magnitude
of 19.1.  This flux level corresponds to the magnitude limit of
the ``low-redshift'' SDSS quasar survey (the $ugri$ selected
quasars which are mostly located at redshifts less than 3.0;
see \citet{richards02}).  Spectra have been obtained for nearly
$20,000$ objects that were not targeted by any of the SDSS selection
algorithms.  In this paper the data set is examined to measure
the completeness of the SDSS quasar survey.  The stellar content
of the completeness survey will be addressed in future papers.
Section\,\ref{observations} describes the spectroscopic survey. The
completeness of the SDSS quasar survey at all stages from imaging
to identified spectra are determined in \S\,\ref{completeness},
and the global survey completeness and quasar surface density 
are calculated.  A discussion of the quasar population not targeted
by the SDSS is given in \S\,\ref{discussion}.


\section{Target Selection and Observations\label{observations}}
%
%
\subsection{The Main SDSS Quasar Survey\label{mainSurvey}}

The SDSS is a project to image of order $10^{4}\,{\rm deg}^{2}$ of sky
mainly in the northern Galactic cap, in five broad photometric bands
($u,g,r,i,z$) to a depth of $r \sim 23$, and to obtain spectra of
$10^{6}$ galaxies and $10^{5}$ quasars selected from the imaging survey
\citep{york00}.  Imaging observations are made with a dedicated
2.5m telescope at Apache Point Observatory in New Mexico, using a
large mosaic CCD camera \citep{gunn98} in a drift-scanning mode.
Absolute astrometry for point sources is accurate to better
than $100$ milliarcseconds rms per coordinate \citep{pier03}.
Site photometricity and extinction monitoring are carried out
simultaneously with a dedicated 20-inch telescope at the observing
site \citep{hogg01}.

The imaging data are reduced and calibrated using the {\tt photo}
software pipeline \citep{lupton01}.  In this study we use the
point-spread function (PSF) magnitudes, which are determined by
convolving the reduced imaging data with a model of the spatial
point-spread function.  The SDSS photometric system is normalized so
that the $u,g,r,i,z$ magnitudes are approximately on the AB system
(Oke \& Gunn 1983; Fukugita et al.\ 1996; Smith et al.\ 2002; c.f.\
discussion by Abazajian et al.\ 2004).  The photometric zeropoint
calibration is accurate to better than $2\%$ (root-mean-squared)
in the $g$, $r$, and $i$ bands, and to better than $3\%$ in the
$u$ and $z$ bands, measured by comparing the photometry of objects
in scan overlap regions.  Throughout this paper we use magnitudes
corrected for Galactic extinction according to \citet{schlegel98}.
Spectroscopic targets are selected by a series of algorithms
\citep[see][]{stoughton02}, and are grouped by three degree
diameter areas or ``tiles'' \citep{blanton03}.  Two fiber-fed double
spectrographs can obtain 640 spectra for each tile; for the main
survey each tile contains 32 sky fibers, and roughly 500 galaxies,
50 quasars, and 50 stars.

Quasar candidates are selected from the SDSS color space
and unresolved matches to sources in the FIRST radio catalog
\citep{becker95}, as described by \citet{richards02}.  These
are the only two categories of quasar candidates which are
specifically targeted for spectroscopic observations in the
SDSS, and the only two we consider for measuring the quasar
survey completeness.  Spectra of quasars are often
obtained because the objects were targeted for spectroscopy
by non-quasar selection algorithms, such as optical matches to
ROSAT sources \citep{stoughton02,anderson03}, various classes of
stars \citep{stoughton02}, or so-called serendipity (SERENDIP)
objects \citep{stoughton02}.  However, those objects are assigned
spectroscopic fibers only if fibers are available after all
candidate quasars, galaxies, sky positions, and calibration stars
corresponding to a given tile have first been assigned fibers.
The contribution of additional quasars from non-quasar selection
algorithms is discussed in \S\,\ref{discussion}.

Objects in the imaging data must first pass a series of image
quality tests.  Then the colors of the objects are compared
to a parameterization of the stellar locus, convolved with the
uncertainties of the colors of the object; objects with colors
outside of the stellar locus region are considered quasar candidates.
Low-redshift ($z \lesssim 3$) ``UV excess'' quasars are selected from
the $ugri$ color cube, while high-redshift quasars are selected from
the $griz$ color cube.  As discussed by \citet{richards02}, there are
additionally some regions of color space inside the stellar locus
from which quasar candidates are explicitly included, and regions
outside the locus for which candidates are excluded.  Specifically,
objects are rejected if they lie inside the regions typically dominated
by white dwarfs, A stars, or M star-white dwarf pairs.  Objects are
included if they lie in the color space expected for quasars with
redshifts between $z=2.5$ and $z=3.0$, which is where the ``quasar
locus'' crosses the stellar locus.  Three additional sets of color
cuts are used to improve the selection of high-redshift ($z>3$)
quasars.  Finally, a simple UV-excess cut is implemented for more
direct comparison with many previous quasar surveys.  All of the
color regions and cuts are given explicitly by \citet{richards02}.
The limiting dereddened magnitude of the survey is $i=19.1$ for
$ugri$ selected (typically low-redshift) and radio matched quasar
candidates.  High-redshift quasar candidates are targeted up to
$i=20.2$, but we consider only those high-redshift candidates with
magnitudes brighter than $19.1$ for this study.  A bright limit of
$i=15.0$ (uncorrected for Galactic extinction) is also imposed to
avoid saturation and crosstalk in the spectroscopic data.

%
%
\subsection{The Point Source Completeness Survey\label{survey}}

To test the performance of the quasar (and other) SDSS target
selection algorithms, a program was undertaken to identify unresolved
sources which were not selected for spectroscopic observations in
the normal course of the survey.  The resulting set of identified
objects can be used to determine the fraction of quasars and other
objects ``missed'' by the main survey, and to determine from their
spectral properties why they were not selected.

The completeness survey is focused on a section of SDSS stripe
number 82 \citep{abazajian03,abazajian04}, which is centered on
the Celestial Equator, and extends from $\alpha_{J2000} = 309\fdg2$
to $59\fdg8$ with a width of just over $2\fdg5$; the area of this
region is $278\,{\rm deg}^{2}$.  The imaging and spectroscopic
observations of stripe 82 are made in the Fall months when the
unobserved regions of the main survey area are not accessible.
All of the imaging runs from which the sources have been selected
for spectroscopic observations are included in the SDSS second and
third data releases \citep{abazajian04, abazajian05}.  The imaging
run numbers are: 2583, 2659, 2662, 2738, 3325, and 3388.  They are
the imaging runs covering the survey area that had the highest
quality at the start of the survey, ensuring that the selection
algorithms would operate under the best available conditions.
For the purposes of the quasar completeness survey, the sky area is
also ideal because it has a high {\em spectroscopic} completeness
(the fraction of main sample quasar candidates with spectroscopic
identifications).

The sample of quasar candidates selected by the algorithm described by
\citet{richards02} from among the objects in the imaging data used
for this survey, differs from the sample targeted for spectroscopy
by the SDSS in the same sky area.  There are two primary reasons
for this.  First, the SDSS spectroscopic targets were selected
by an earlier version of the quasar algorithm.  Most notably,
there are now more high-$z$ candidates --- those selected by
their location in the $griz$ color cube \citep{richards02} ---
which also causes the set of unobserved targets to be dominated
by high-$z$ candidates. Second, variations in object brightness,
due to measurement uncertainty as well as real variability, cause
the population of quasars to change slightly from one imaging run
to the next.  Studies of the variability of SDSS quasars show that
the $i$ band rms magnitude differences are $\approx 0.1$ on the
timescale of a year \citep{vandenberk04,ivezic03}; this can cause
some quasars to either make or miss the magnitude limit from one run
to the next.  However, over $90\%$ of the quasar candidates selected
for this program by the final algorithm were previously selected in
other runs covering the same area of sky using an earlier version
of the algorithm.

The goal of this program is to measure the angular density
of quasars on the sky which are missed by the final selection
algorithm.  This is accomplished by spectroscopically identifying
objects selected from all regions of color space.  The potential
targets for this program were selected as unresolved (point)
sources, identified by the {\tt photo} pipeline, in order to avoid
targeting an excessive number of galaxies.  The survey completeness
for quasars with extended image profiles (mainly lower redshift
active galactic nuclei) is addressed in \S\,\ref{extended} using
the main SDSS galaxy sample. (For other discussions of samples
of SDSS active galactic nuclei with extended image profiles, see
\citet{richards02,strauss02,kauffmann03}, and \citet{hao04}.)
Unresolved sources were selected which have magnitudes fainter
than $i = 15.0$ (uncorrected for Galactic extinction) and brighter
than dereddened $i = 19.1$; these limits are adopted in order to
avoid saturation at the bright end and to match the quasar survey
limit at the faint end.  Additional constraints were imposed on the
quality of the images (e.g. bad CCD columns were avoided) as they
were for the main quasar survey, as described by \citet{richards02}
and \citet{stoughton02}.  No color cuts were applied.  The selection
process up to this stage is identical to that used by the main
SDSS quasar survey before the color selection algorithm is applied,
except that here only point sources were selected.

The total number of objects selected as potential follow-up targets
(regardless of color) in the completeness program area is $674,842$.
Of those, 3192 were targeted as quasar candidates by the final
SDSS quasar algorithm (including only color and radio matching
selection).  Of the targeted quasar candidates, 2909 ($91.1\%$)
have been spectroscopically observed, mostly as part of the main
SDSS quasar survey.  Some of the remaining candidates have been
targeted for spectroscopy in the quasar survey, but have simply not
yet been observed.  Other candidates have not been targeted for
the spectroscopic quasar survey for reasons discussed earlier in
this section.  A small number of objects in the sample are quasars
that were not selected by the final algorithm, but instead by
other targeting algorithms such as X-ray source matching or star
candidate selection; these are discussed in \S\,\ref{discussion}.
The number of objects in the program area {\em not} selected
as quasar candidates is prohibitively large for a comprehensive
spectroscopic survey, therefore the set of objects was sparsely
sampled for spectroscopy using the criteria described next.

\begin{figure} 
  \plotone{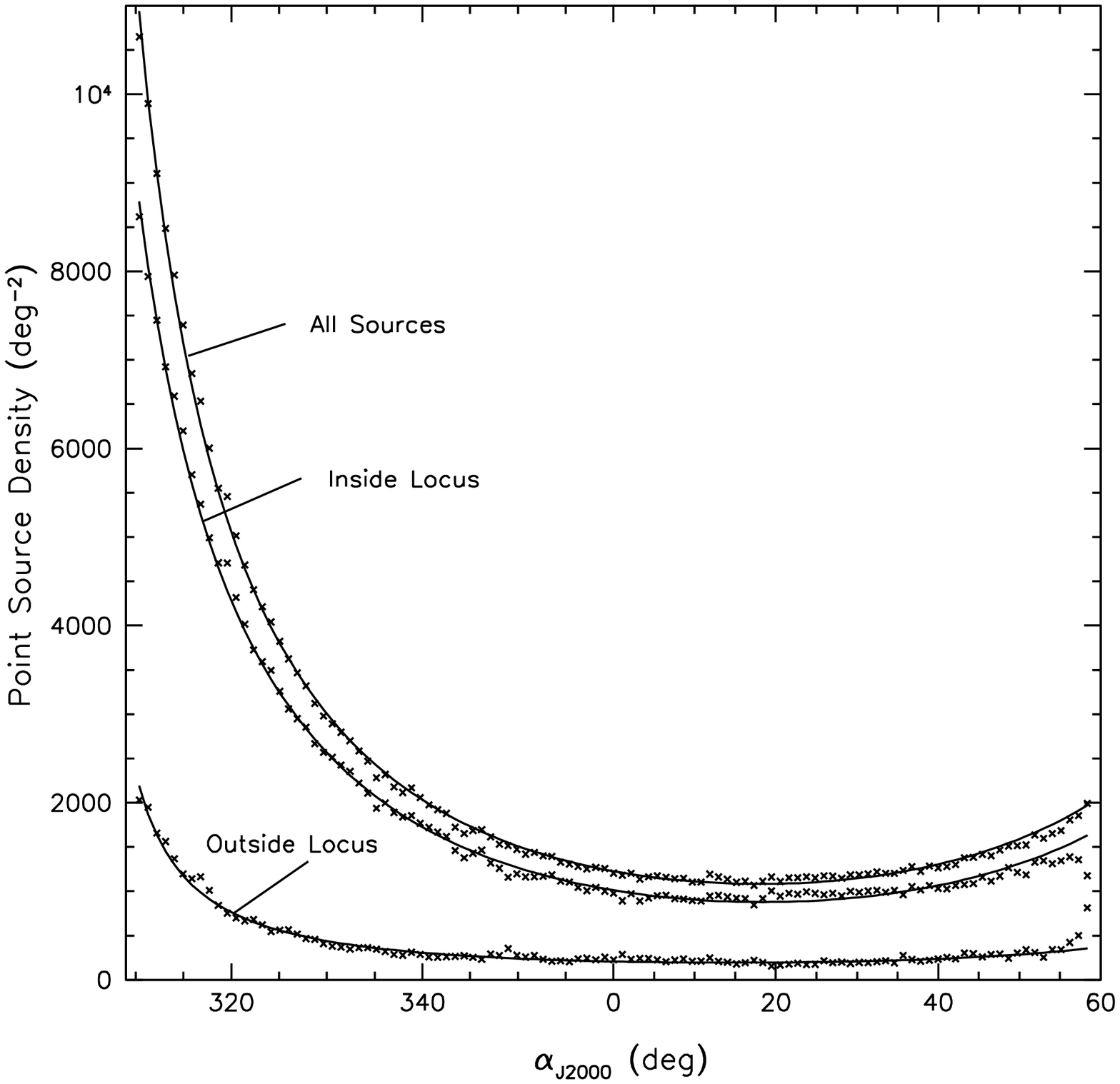} 
  \caption{Density of non-quasar-candidate point sources to $i=19.1$ as a
  function of right ascension for a 2.5\,deg wide stripe centered on the
  Celestial equator.  The density of all sources is shown by the
  top set of points, while the densities of sources defined to be
  ``inside'' and ``outside'' the stellar locus are shown by the
  middle and bottom sets of points respectively.  Model fits to the
  densities (see text) are shown as solid lines.
  \label{densityPlot}}
\end{figure}

The density of point sources is a strong function of right
ascension, as shown in Figure\,\ref{densityPlot}; this is not
surprising given the large range in Galactic latitude ($-60\fdg6 <
b < -23\fdg7$) covered by the SDSS stripe.  There is very little
change in density across the $2\fdg5$ of declination.  To provide a
more uniform distribution of targets for the spectroscopic plates,
the probability of selecting an object was varied inversely as the
density at its right ascension.  The density of point sources was
fit with a fourth-order polynomial, and the selection probability was
proportional to the inverse of the fit.

In practice, the selection probability was set to 100 times the 
inverse of the point source density, and an object was selected if
the decimal part of 10 times its right ascension (in degrees) was less than
the selection probability.  The sampling rate thus ranged from
nearly unity at the minimum of the point source density, to as low
as $10\%$ at the high density end of the stripe.

A simpler parameterization of the point source density,
which also provides a better fit to the data, is a modified secant function
\begin{eqnarray}
  n_{pt}(\alpha_{J2000}) = a_{0} + a_{1}\sec(a_{2}\alpha_{J2000} + a_{3}),
  \label{Eq_secant}
\end{eqnarray}
where the $a_{i}$ are constants.  (This function was not used for
the sparse sampling since it was first fit after the sampling
was completed.)  For the entire set of non-quasar candidate
point sources, the values are $a_{0}=-1647.8$\,deg$^{-2}$,
$a_{1}=2730.8$\,deg$^{-2}$, $a_{2}=1.097$, and $a_{3}=-23\fdg17$;
the fit is shown in Fig.\,\ref{densityPlot}.  The density of all
sources is important for determining the density of untargeted
quasars (\S\,\ref{misseddensity}).

After sparse sampling the sources based on density, non-quasar
candidate targets for spectroscopic follow-up were selected.
Observations were carried out in two seasons, Fall 2002 and Fall
2003, and the spectroscopic targets were selected in different
ways each season.  First, before the Fall 2002 observations,
objects were selected at random from any point in color space.
This is the simplest method, but may not be the most efficient
method of investigating quasar completeness.  In the densest parts
of the stellar locus, the ratio of stars to quasars is maximal, so
it is most efficient to avoid those regions.   Because of the high
precision of the SDSS photometry, the core of the stellar locus is
very narrow and fills a relatively small volume of color space (there
is no reason to expect the color distribution of quasars to mimic
this stellar locus core).

Two other selection methods --- used before the Fall 2003 observations
--- were designed to sample the volume of color space outside of the
core of the stellar locus more often than it would be for purely
random color selection.  To implement these options, a pair of
lines was fit to the stellar locus in the $gri$ color plane, and
the corresponding standard deviation of the color distance from
the lines was determined.  The color distances from the lines were
calculated for all of the objects.  Those objects more than one
standard deviation away from the locus fit were deemed ``outside''
of the stellar locus, and spectroscopic targets were drawn randomly
from this set.  Objects closer than one standard deviation ---
defined to be ``inside'' the stellar locus --- were given a priority
weight roughly proportional to the distance from the locus line fits.
Spectroscopic targets from this sample were selected at random but
weighted by the priority.  The numbers of spectroscopic targets
in each sample were selected to be roughly equal, which ensured
that more targets outside the locus would be observed than for a
strictly random color space selection.  The densities of all point
sources inside and outside the stellar locus core as a function of
$\alpha_{J2000}$ are shown in Fig.\,\ref{densityPlot}.  The object
selection used here is much simpler than the SDSS quasar selection
\citep{richards02}, and is designed only to select a large fraction
of targets which are close to, but not inside, the densest parts
of the stellar locus.

\begin{figure}
  \plotone{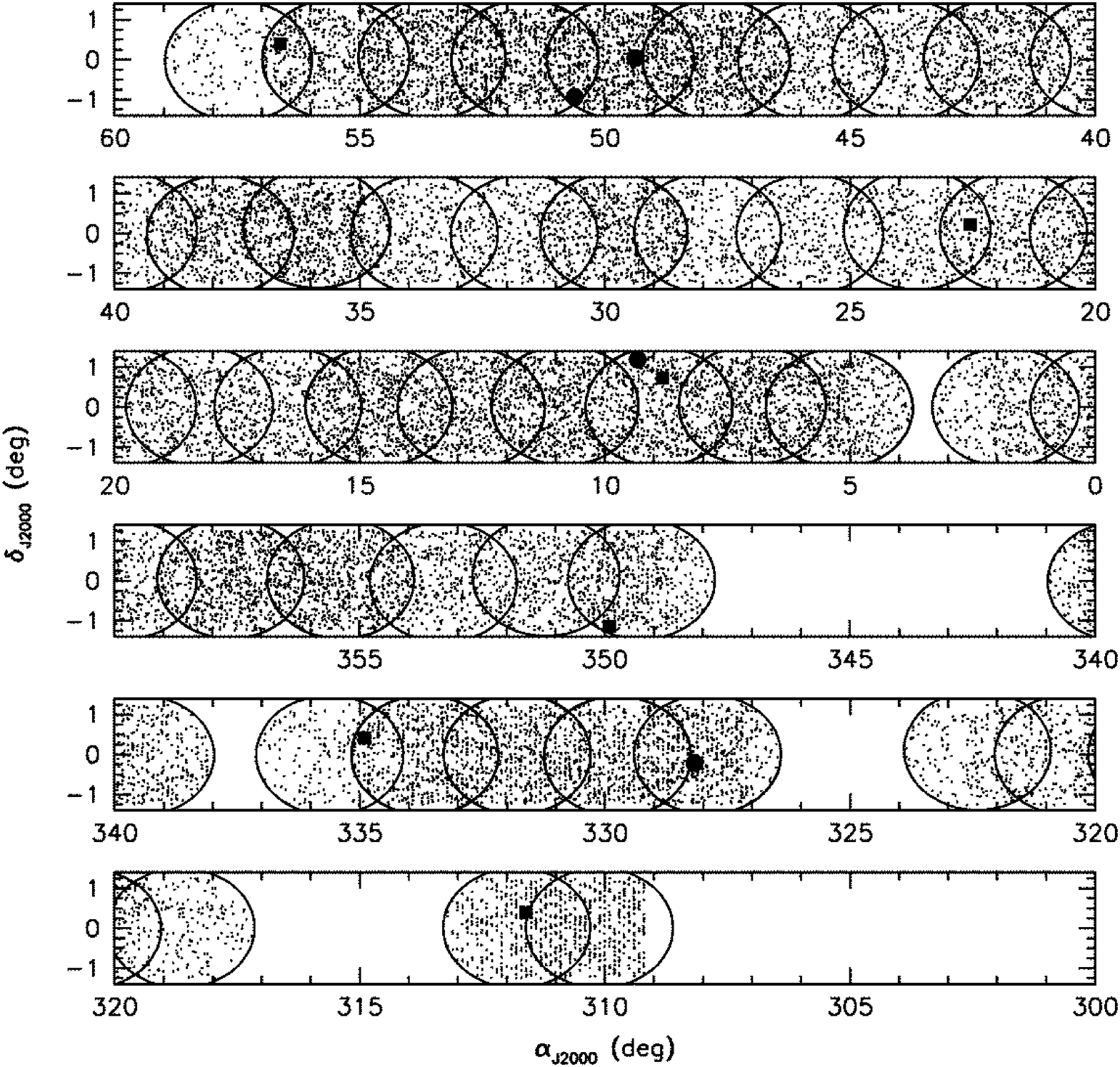}
  \caption{Distribution on the sky of the objects spectroscopically
  observed for this study.  The density of targets is roughly
  constant with $\alpha_{J2000}$ due to sparse sampling, but not all
  of the targets have been observed.  The ten quasars are plotted
  as filled squares if selected from the random color sample,
  and as filled circles if selected as outside the stellar locus.
  The spectroscopic plate boundaries are shown as large circles.
  \label{coordPlot}}
\end{figure}

Two sets of spectroscopic plates were designed.  The first set
was designed before the Fall 2002 observations for the objects in
the completely random target selection sample. The second set was
designed before the Fall 2003 observations for the objects selected
to be either inside or outside the core of the stellar locus.
There were 41 plates observed for the randomly selected objects, and
26 for those selected by locus distance.  Figure\,\ref{coordPlot}
shows the distribution on the sky of the 67 plates used in this
study.  The plates cover an area of $233.2\,{\rm deg}^2$ within the
$278\,{\rm deg}^2$ image stripe area.  The plate areas often overlap,
but different sets of objects are observed.  The sparse sampling
procedure described above causes the targets to be drawn from
narrow ranges of $\alpha_{J2000}$ where the stellar density is high
(see Fig.\,\ref{coordPlot}), but this has no effect on the survey.
The fibers corresponding to each plate configuration were shared
with other programs, but this does not affect the results here;
about $50\%$ of the available science fibers were allocated to the
completeness program.  In total, 12494 randomly selected targets
were observed spectroscopically, while 3946 targets inside, and 3090
outside the stellar locus were observed.  Spectroscopic observations
and data processing were performed in the same manner as is done
for the main SDSS survey \citep{stoughton02}.

\subsection{Spectroscopic Results\label{spectro}}
The reduced and calibrated spectra of the 19530 objects were given
preliminary identifications by the {\tt Spectro1d} pipeline.
The pipeline identifications and measurements are not perfect,
however, and often fail when the spectra have unusual properties,
such as may be the case for a significant number of the quasars
found in a survey like this one.  Therefore, the spectra of all
19530 observed objects were manually inspected by one author, and
those with unusual or ambiguous identifications were inspected by
at least five other authors.

In the entire spectroscopic sample, only ten objects were identified
as quasars (all but one were correctly identified by the {\tt
Spectro1d} pipeline).  Of these, six were found among the randomly
selected targets, four were from the sample outside the stellar
locus, and none were found from the sample inside the stellar locus.
Four have redshifts between 2.7 and 3.0.  The fact that so few
quasars were found among objects not selected as quasar candidates
suggests that the selection completeness is high; the completeness
will be quantified in the next section.

\begin{figure}
  \plotone{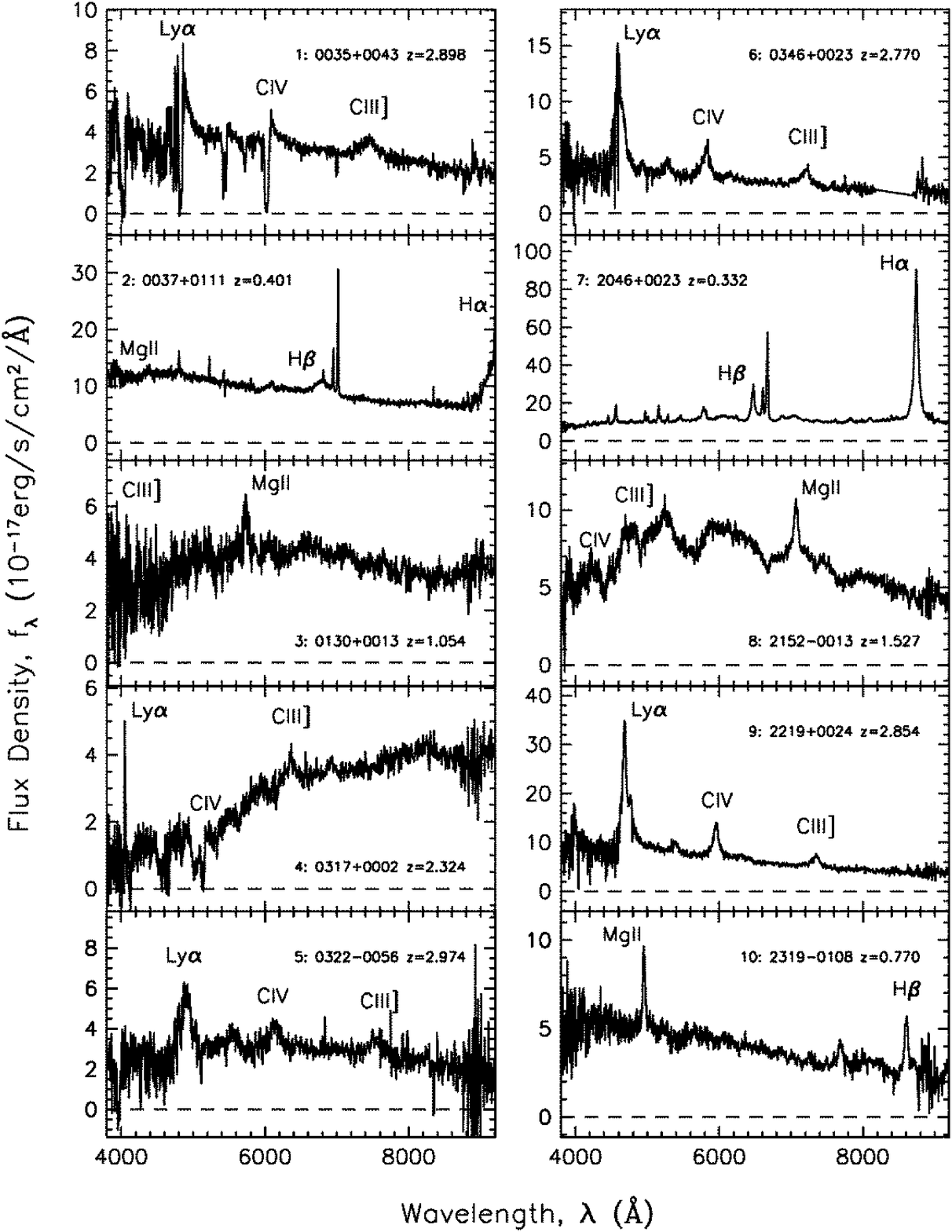}
  \caption{Spectra of quasars, ordered by $\alpha_{J2000}$, that were
  not targeted as quasar candidates by the SDSS quasar selection
  algorithm.  Identifications are given by the four digit truncated
  coordinates, and the index number listed in Table\,\ref{tab1}.
  The locations of several major emission lines are labeled.
  The spectra have been smoothed by a five-pixel boxcar for
  clarity. The spectral resolution $\lambda/\Delta\lambda$, is
  approximately 1200 after smoothing.
  \label{specPlot}}
\end{figure}

The spectra of the ten quasars are shown in Fig.\,\ref{specPlot},
and Table\,\ref{tab1} gives a summary of their properties.  All of
the quasars are newly discovered by this program. Color-color and
color-magnitude diagrams for all of the observed objects are shown in
Fig.\,\ref{colorcolorPlot}, with the quasar locations highlighted.
The colors of the ten quasars as a function of redshift, in
comparison with the median color vs.\ redshift curves for the
2191 quasars in the area selected using the final SDSS algorithm,
are shown in Fig.\,\ref{zColorPlot}.  A discussion of each of the
quasars is given in \S\,\ref{notes}.

\section{The SDSS Quasar Survey Completeness\label{completeness}}

We first discuss why the quasars found in the completeness survey
were not targeted by the SDSS quasar algorithm.  Next, we measure
the densities on the sky of targeted and missed quasars, and
use these results to determine the completeness of the quasar
selection algorithm.  Incompleteness due to image defects and
unidentifiable spectra are also discussed, and an estimate of
the completeness for extended sources is also made.  Finally,
the end-to-end completeness of the of the SDSS quasar survey is
calculated, along with the corresponding estimate of the true
density of quasars to the limiting magnitude.

%
\subsection{Notes on Individual Objects\label{notes}}

The reasons each of the quasars in the completeness survey were
missed by the SDSS quasar selection algorithm are given here.
No quasar was rejected because it is located in the white dwarf,
A star, or white dwarf + M star ``exclusion boxes'' --- regions
of color space outside the core of the stellar locus, in which
the density of certain types of stars would cause a significant
drop in quasar selection efficiency \citep{richards02}.  All of the
quasars have colors consistent with the stellar locus in the $ugri$
color cube.  The spectroscopic and imaging magnitudes are consistent
for all but two of the objects, both of which appear to have varied
between the imaging and spectroscopic epochs.

\begin{figure}
  \plotone{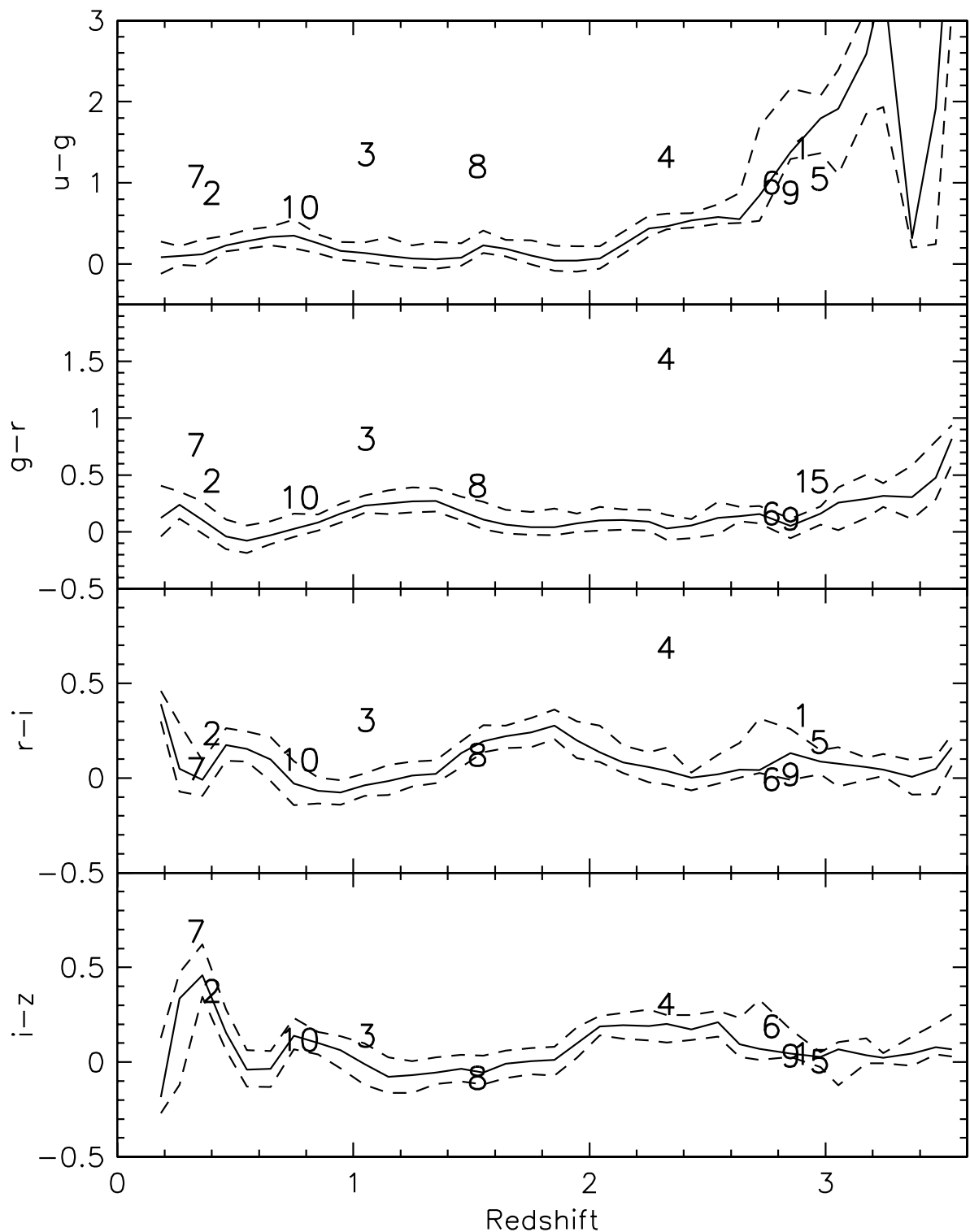}
  \caption{Color as a function of redshift for the ten quasars
  discovered in the quasar completeness program.  The numerals
  correspond to the list in Table\,\ref{tab1}.  The solid lines
  show the median colors as a function of redshift for the 2191
  quasars selected by the SDSS quasar algorithm.  The dashed
  lines bound the $68.3\%$ (nominal 1$\sigma$) confidence intervals
  about the median lines.
  \label{zColorPlot}}
\end{figure}

{\bf 1.~SDSS J003517.95+004333.7} ($z=2.898$): This is a
high-ionization broad absorption line quasar (BALQSO), with BAL
troughs corresponding to C\,{\sc iv}, Si\,{\sc iv}, N\,{\sc v},
and Ly\,$\alpha$ transitions.  The quasar is somewhat redder in
the $g-r$ and $r-i$ colors than are most quasars at the same redshift
(Fig.\,\ref{zColorPlot}), and it is closer to the stellar locus
because of this.  BALQSOs tend to be significantly redder than
non-BALQSOs \citep[e.g.][and references therein]{reichard03,brotherton01},
but it is not the broad absorption lines themselves that cause the
color shift.

{\bf 2.~SDSS J003719.85+011114.6} ($z=0.401$): This quasar is a
stellar locus outlier in the $griz$ color cube, but was rejected
as a candidate because its colors are located in a region
which is excluded in order to avoid overwhelming the high-$z$
candidate samples with low-$z$ quasars \citep{richards02}.
Quasar SDSS~J204626.11+002337.7 was rejected for the same reason.
The fact that both are low-$z$ quasars demonstrates that the cuts
do what they were designed to do.  What is more important is that
neither quasar was selected by the low-$z$ $ugri$ color algorithm.
Relative to other quasars at the same redshift, this quasar is
significantly redder in the $u-g$ and $g-r$ colors, which places
it much closer to the stellar locus.  The spectral slope is
redder (flatter) than that of the average quasar at wavelengths
shortward of about $4000${\AA}.  A power-law fit to this quasar
at wavelengths near $3060${\AA} and $4200${\AA} has a slope of
$\alpha_{\lambda}=-0.94$, compared to $\alpha_{\lambda}=-1.80$
for a composite quasar spectrum \citep{vandenberk01} at the same
wavelengths.  There are no significant stellar absorption features,
except possibly H\,8 at $\lambda=5443${\AA}, which means that a host
galaxy component is unlikely to make the quasar colors appear redder.
The quasar is likely to be either reddened by dust, or to have an
intrinsically redder than average continuum \citep{richards03}.

The spectroscopic $g$, $r$, and $i$ magnitudes --- obtained by
convolving the filter transmission curves with the spectrum (the
$u$ and $z$ bands are not covered by the SDSS spectra) --- have
changed significantly from the image magnitudes.  The spectroscopic
magnitudes are brighter by $0.36, 0.28$, and $0.17$ in the $g$, $r$,
and $i$ bands respectively, and thus the quasar has also become
bluer, which is consistent with the known variability properties of
quasars \citep[e.g.][]{vandenberk04}.  There are no obvious problems
with either the image or spectrum of this object, which implies
that the variability is real.  Given the relatively small $g-r$
and $r-i$ color changes, it is unlikely that the quasar would have
been selected as a candidate at the spectroscopic epoch, however,
without the $u$ magnitude, that cannot be determined with certainty.

{\bf 3.~SDSS J013011.42+001314.6} ($z=1.054$): This quasar
is significantly redder than the average quasar in all four
colors, and is progressively redder toward shorter wavelength bands
(Fig.\,\ref{zColorPlot}).  The spectrum shows that the continuum
appears to rise toward longer wavelengths, opposite to what is
normally observed for quasars at this redshift.  There is no
obvious BAL feature in the spectrum, but the S/N is too low to
allow detection of any but the strongest absorption features.
The colors relative to other quasars indicate that the quasar is
probably dust reddened \citep{richards03}.

{\bf 4.~SDSS J031732.20+000209.7} ($z=2.324$): This is a highly
reddened broad absorption line quasar.  Large magnitude uncertainties
(due to little flux) in the $u$ and $g$ bands place the colors
of this quasar inside the stellar locus region.  The redshift is
based on the C\,{\sc iii}]$\lambda 1909$ emission line which is not
absorbed.  There is BAL absorption from C\,{\sc iv}, Si\,{\sc iv},
N\,{\sc v}, and Ly\,$\alpha$ transitions.  There may be a weaker
absorption feature corresponding to A\,{\sc iii}, which would make
this a low-ionization BALQSO.  Low-ionization BALQSOs tend to be
significantly redder than non-BALQSOs \citep[e.g.][and references
therein]{reichard03,brotherton01}.  There is a strong narrow
Ly\,$\alpha$ emission line feature, which may either arise from
narrow-line gas outside of the absorbing BAL material, or may be
the remaining portion of the broad Ly\,$\alpha$ emission line
that is not absorbed by Ly\,$\alpha$ and N\,{\sc v} gas.

\begin{figure}
  \plotone{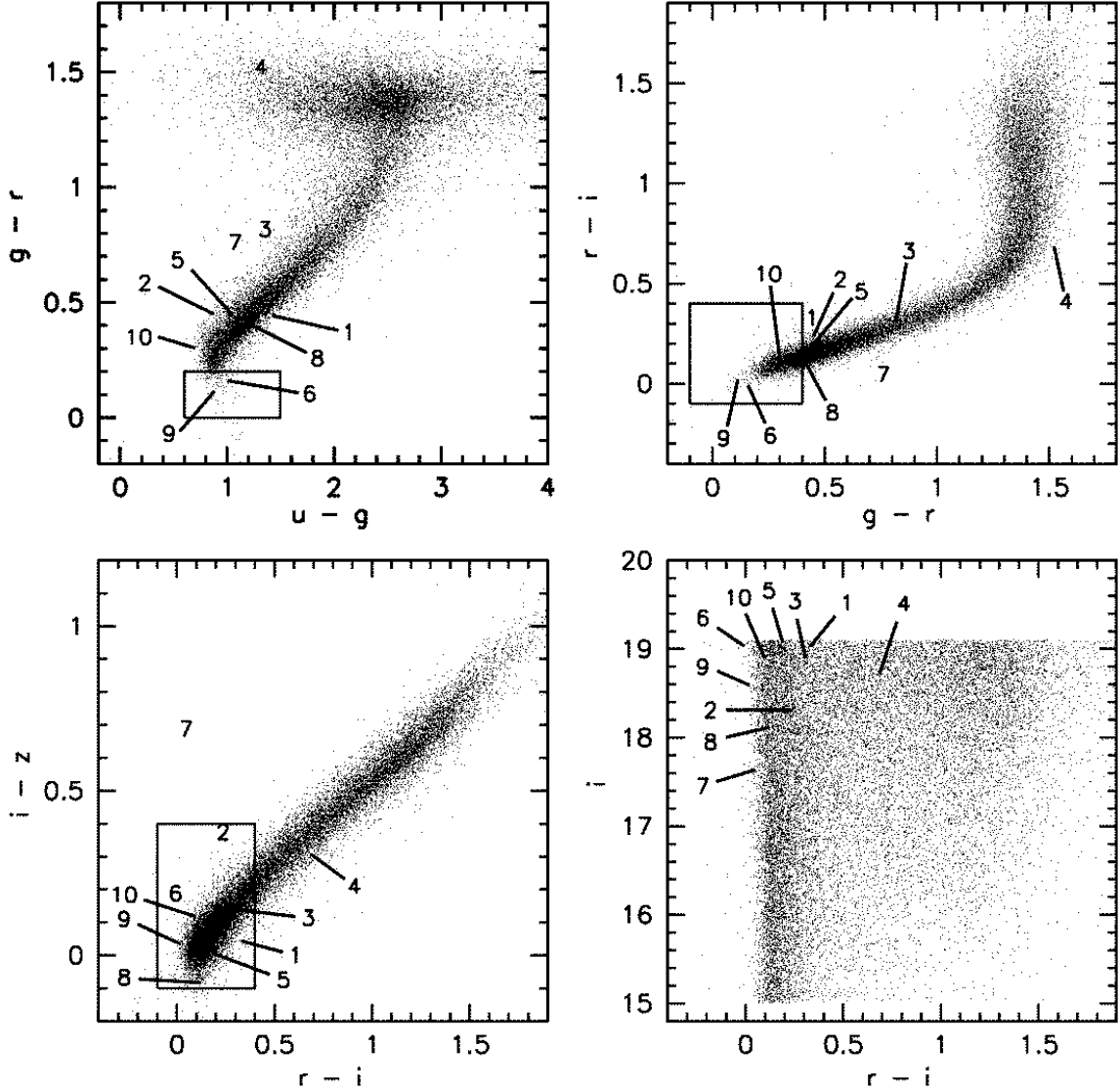}
  \caption{Color-color and color-magnitude diagrams for the 19530 objects
  spectroscopically observed for this program.  The locations of the
  identified quasars are shown as numerals corresponding to the 
  RA-sorted list in Table\,\ref{tab1}.  The region of
  color space containing part of the stellar locus, but which is
  sparsely sampled by the SDSS quasar algorithm --- the ``mid-$z$
  inclusion region'' described by \citet{richards02} --- is indicated
  by rectangles.  Two of the quasars, J$0346$+$0023$ (number 6) and
  J$2219$+$0024$ (number 9), lie inside this mid-$z$ inclusion
  region.
  \label{colorcolorPlot}}
\end{figure}

{\bf 5.~SDSS J032228.99-005628.6} ($z=2.974$): This quasar
has the highest redshift in the sample; redshifts of three of
the other quasars (J003517.95+004333.7,\\ J034629.02+002337.7,
J221936.37+002434.1) differ by less than $0.21$.  Quasars at
this redshift often have colors very similar to stars in the
stellar locus.  Because of this, the selection algorithm allows
some objects to be randomly targeted which lie inside a ``mid-$z$''
inclusion region, shown by the boxes in Fig.\,\ref{colorcolorPlot}.
The colors of this quasar would place it inside the inclusion region,
except for the $g-r$ color which is redder than typical for the
quasar redshift.  There is a strong intervening Mg\,{\sc ii} system
in the spectrum at $z=0.8243$.  Gas and dust associated with this
system may redden the quasar spectrum, but probably not in such a
way as to redden only the $g-r$ color.  It seems more likely that
the color is due to the intrinsic properties of the quasar itself.

{\bf 6.~SDSS J034629.02+002337.7} ($z=2.770$): This quasar could
have been selected as a candidate because it is located in the
mid-$z$ inclusion box (Fig.\,\ref{colorcolorPlot}), and is outside
of the 2$\sigma$ stellar locus region \citep[see][]{richards02}.
However, it was not among the randomly sampled mid-$z$ objects.
The colors are normal for a quasar at its redshift.

{\bf 7.~SDSS J204626.11+002337.7} ($z=0.332$): This quasar has the
lowest redshift in the sample.  It is a stellar locus outlier in
the $griz$ color cube, but was rejected because its colors are also
consistent with being a low-$z$ object.  This is the same reason
quasar SDSS~J003719.85+011114.6 was rejected.  It was also not
selected by the low-$z$ $ugri$ color algorithm.  Relative to other
quasars with the same redshift, this quasar is redder in all but the
$r-i$ color (Fig.\,\ref{zColorPlot}).  There is a possible Ca\,{\sc
II}$\lambda 3934$ stellar absorption feature at $\lambda=5242${\AA},
which may indicate that host galaxy light contributes significantly
and causes the total spectrum to be redder than the quasar component.

{\bf 8.~SDSS J215241.89-001308.7} ($z=1.527$): This quasar is a
probable low-ionization BALQSO, with absorption troughs due to
Al\,{\sc iii}, Al\,{\sc ii}, and Mg\,{\sc ii} at $z\approx1.385$.
The C\,{\sc iii}] line is relatively weak, while the UV Fe
complexes are strong relative to other quasars.  There is also a
possible complex narrow C\,{\sc iv} absorption system at $z=1.4855$.
The $u-g$ color is significantly redder than other quasars at this
redshift, which is the primary reason the colors are located inside
the stellar locus, and the quasar was not selected as a candidate.

{\bf 9.~SDSS J221936.37+002434.1} ($z=2.854$): Although
this quasar has colors inside the mid-$z$ inclusion region
(Fig.\,\ref{colorcolorPlot}), it was rejected as a candidate because
it is within the $2\sigma$ region surrounding the stellar locus.
A $4\sigma$ region surrounding the locus is normally used for quasar
selection, but the smaller $2\sigma$ region is used when considering
the mid-$z$ color space \citep{richards02}.  The colors are normal
for a quasar at its redshift.

The spectroscopic $g$, $r$, and $i$ magnitudes are significantly
brighter than the imaging magnitudes by $0.22$, $0.16$, and $0.14$
magnitudes respectively, and thus the quasar is also slightly bluer
at the spectroscopic epoch.  There is no evidence from the quasar
image or spectrum that the variability is not real.  The small
color changes are unlikely to have changed the quasar candidate
selection outcome, but as with SDSS J003719.85+011114.6, that cannot
be determined with certainty without the $u$ band magnitude at the
spectroscopic epoch.

{\bf 10.~SDSS J231937.64-010836.1} ($z=0.770$): This quasar
is marginally redder in the $u-g$ and $g-r$ colors than typical
quasars at the same redshift.  It is mainly the $u-g$ color that
places the quasar colors too close to the stellar locus relative
to other quasars.  There is no indication from the spectrum that
the quasar is unusual in any other way, except perhaps that the
Mg\,{\sc ii} emission line is relatively narrow.

\subsection{Density of Targeted Quasars\label{quasardensity}}

The completeness of the quasar selection algorithm can be determined
by comparing the density on the sky of quasars which are and are
not targeted as quasar candidates.  The density of targeted quasars
is simply the number of such quasars which are spectroscopically
verified, divided by the area of sky from which they were discovered,
corrected for the number of quasars expected among the candidates
which were not spectroscopically observed.

In the imaging data of the completeness program survey area, there
are 3192 objects targeted by the final SDSS quasar algorithm as
candidates.  Of these, 2909 were spectroscopically observed as part
of the SDSS, and 2189 were verified to be quasars.  The non-quasars
are essentially all stars, except that there may be some unidentified
BL\,Lac objects \citep{anderson03,collinge04}, which would not be
counted among the broad-line quasar sample in any case.  All of the
quasar candidate spectra were manually inspected by at least two
of the authors.  A large fraction of the survey area is included in
the SDSS data releases \citep{abazajian03,abazajian04,abazajian05},
and 1824 of the quasars are part
of the DR1 quasar sample \citep{schneider03}. Three additional
candidates were matched with objects in the NED\footnote{The
NASA/IPAC Extragalactic Database (NED) is operated by the Jet
Propulsion Laboratory, California Institute of Technology, under
contract with the National Aeronautics and Space Administration.}
database; two are quasars and one is a BL\,Lac with no measurable
redshift.  The two NED quasars can be added to our list of verified
quasars to bring the total to 2191.  The efficiency of the algorithm
(the fraction of candidates that are verified quasars) is $75.3\%$.
This number is higher than the overall efficiency of $66.0\%$
reported by \citet{richards02}, but consistent with the efficiency
of $75.0\%$ they found for the low-$z$ selected sample.  This is
expected since our sample more closely resembles the low-$z$ SDSS
sample; the magnitude limits (brighter than the high-$z$ limit)
are the same and most of the fainter high-$z$ candidates --- which
have a substantially lower efficiency ($54.4\%$) --- are excluded
from our sample.

\begin{figure}
  \plotone{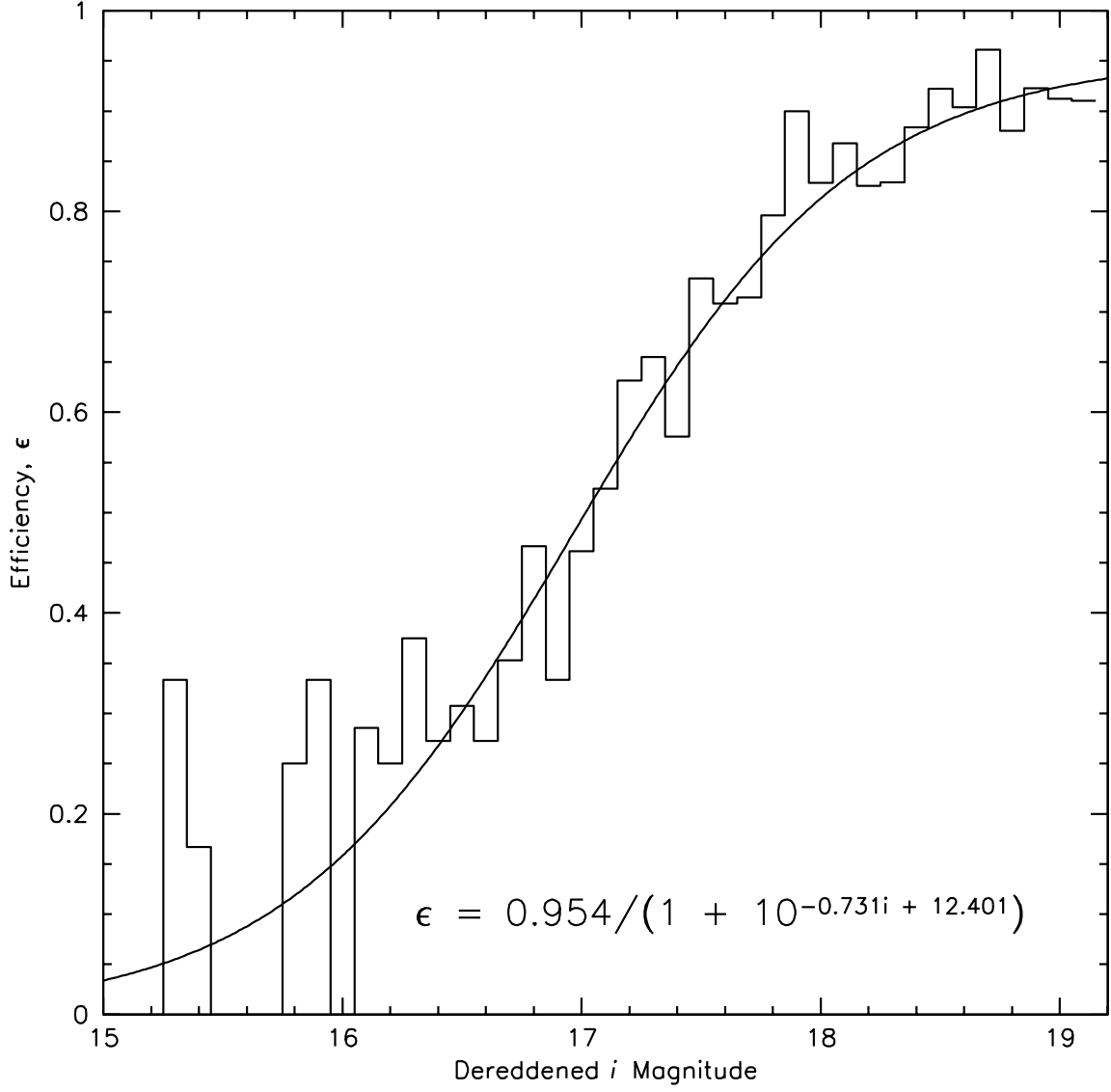}
  \caption{Quasar selection efficiency as a function of the 
  dereddened $i$ magnitude.  The fit to the data, given by
  eq.\,(\ref{magEffEq}), is shown with a solid line.
  \label{magEffPlot}}
\end{figure}

This overall efficiency could be used directly to estimate the number
of quasars expected among the remaining 280 unobserved candidates.
However, the efficiency is a strong function of apparent magnitude,
and depends upon the type of candidate (high-$z$ or low-$z$).  We must
account for possible differences in the distribution of types of
candidates and the magnitude distributions of the observed and
unobserved samples.  Figure\,\ref{magEffPlot} shows the low-$z$
selection efficiency $\epsilon$, as a function of dereddened $i$
magnitude, for the sample of 2912 spectroscopically identified
candidates.   The efficiency change is mainly a reflection of the
varying densities of stars and quasars with magnitude.  The efficiency
as a function of magnitude is well fit by the equation
\begin{eqnarray}
  \epsilon(i) = 0.954/(1 + 10^{-0.731i + 12.401}),\label{magEffEq}
\end{eqnarray}
which is the form expected if the density of targeted stars and
quasars both increase roughly exponentially with magnitude, but at
different rates.  Summing the values of the efficiencies at the
magnitudes of the low-$z$ candidates gives 89 expected additional
quasars.

More than half of the 280 unobserved candidates were selected only
by the high-$z$ color algorithm, which is a much higher fraction
than for all of the candidates, as explained in \S\,\ref{survey}.
There are too few verified quasars from the high-$z$-only sample
to reliably determine the efficiency as a function of magnitude.
The number of expected quasars from the remaining high-$z$ sample
was determined using only a single global efficiency value.
The estimated number of additional high-$z$-only quasars is 18.

The estimated total number of unresolved selected quasars in
the survey area is the sum of the verified quasars and those
expected among the candidates.  In the program survey area of
$277.6$\,deg$^2$, the estimated total number of selected quasars
is 2298.  Assuming a binomial distribution of quasar counts,
the density of selected quasars is $8.28\pm0.14$\,deg$^{-2}$ at
the $90\%$ confidence level.  This is lower than the density of
$\approx 10$\,deg$^{-1}$ found by \citet{richards02} in the low-$z$
SDSS sample, but the difference can be partly accounted for by the
omission of extended sources in our sample (see \S\,\ref{extended}).

\subsection{Density of Missed Quasars\label{misseddensity}}

The density of quasars missed by the SDSS selection algorithm can
be found by comparing the number of detected quasars to the number
expected, given the probability that each spectroscopically observed
object is a quasar.  That is, the total number of quasars expected in
a survey $N_{q,exp}$, is the sum of the probabilities $P_{q,i}$, that
each object is a quasar
\begin{eqnarray} 
  N_{q,exp} & = & \sum P_{q,i}.
\end{eqnarray}

The probability that an object chosen at random from a sample
containing both stars and quasars is actually a quasar, is equal
to the ratio of the density of quasars $n_{q}$, to the density of
all objects $n_{obj}$, at the location of the object in both position
and color space
\begin{eqnarray}
  P_{q} = n_{q}/n_{obj} = n_{q}/(n_{q} + n_{s}),
\end{eqnarray}
where $n_{s}$ is the density of stars.  We assume that the quasar
density is constant over the survey area.  This should be a
good approximation, since the clustering two-point correlation
function is negligibly small on the scale of the survey area
\citep[e.g.][]{croom04}.  As shown in Fig.\,\ref{densityPlot}
and discussed in \S\ref{survey}, the density of unresolved objects
varies strongly as a function of right ascension.  The probability
that a selected object is a quasar is therefore also a function
of right ascension.  The density is also different for each
subsample of objects (random color, inside locus, outside locus),
but in each case it can be fit with the same functional form as
equation\,\ref{Eq_secant}.  The parameter values for the fits to the
density as a function of $\alpha_{J2000}$ are $a_{i}:\{-1654.0\,{\rm
deg}^{-2}, 2532.7\,{\rm deg}^{-2}, 1.070, -23\fdg00\}$ for objects
selected to be inside the stellar locus, and $a_{i}:\{-108.7\,{\rm
deg}^{-2}, 302.9\,{\rm deg}^{-2}, 1.219, -22\fdg09\}$ for objects
outside the stellar locus.

For the collection of spectroscopically observed objects, the total
number of quasars expected is then
\begin{eqnarray} 
  N_{q,exp} & = & \sum n_{q,i}/(n_{q,i} + n_{s,i}) \\
            & = & n_{q} \sum 1/n_{obj,i}.  
\end{eqnarray} 
where again it is assumed that the quasar density is a constant.
The density of missed quasars $n_{q,missed}$, is then simply
\begin{eqnarray} 
  n_{q,missed} = N_{q,obs}/  \sum 1/n_{obj,i}.
  \label{Eq_nqexp}
\end{eqnarray}
where the number of expected quasars is set equal to the number of
observed quasars, $N_{q,obs}$.

The total number of quasars found in the completeness survey is ten.
Six were detected in the random color sample, four were detected
from the sample outside the stellar locus, and none were found
from the sample inside the stellar locus.  Using these numbers
and the object density at the locations of all of the observed
targets, the density of quasars missed by the SDSS selection
algorithm is $n_{q} = 0.44^{+0.31}_{-0.17}\,{\rm deg}^{-2}$.
The upper and lower uncertainties bound the $90\%$ confidence
interval, assuming the number of detected quasars follows a binomial
probability distribution.  The small number of objects prevents
us from determining the density as a function of color, but the
densities calculated for each of the subsamples are consistent with
(but less strongly constrained than) the global density value:
$0.70^{+0.68}_{-0.31}\,{\rm deg}^{-2}$ for the random color selected
sample, $0.37^{+0.48}_{-0.19}\,{\rm deg}^{-2}$ for the sample outside
the stellar locus, and less than $0.71\,{\rm deg}^{-2}$ at the $90\%$
confidence level for the sample inside the stellar locus.

\subsection{Completeness of the SDSS Quasar Selection Algorithm
  \label{algorithm}}

The completeness $C_q$, of the selection algorithm is the density of quasars
selected by the SDSS algorithm $n_{q,selected}$, to the density of all
quasars $n_{q,total}$
\begin{eqnarray}
  C_{q} & = & n_{q,selected}/n_{q,total} \\
        & = & n_{q,selected}/(n_{q,selected} + n_{q,missed}),
  \label{Eq_complete}
\end{eqnarray}
where $n_{q,missed}$ is the density of quasars missed by the
SDSS algorithm.  Using the numbers in \S\S\ref{quasardensity} and
\ref{misseddensity}, the completeness of the SDSS quasar selection
algorithm is $C_q = 94.9^{+2.6}_{-3.8}\%$ at the $90\%$ confidence
level. 

The completeness value of nearly $95\%$ is well
above the initial survey goal of $90\%$\footnote{See:
http://www-sdss.fnal.gov:8000/edoc/requirements/scireq/scireq.html},
while maintaining a very high efficiency ($\approx75\%$,
\S\ref{quasardensity}).  As a comparison, the estimated completeness
of combined variability and proper motion selected quasar samples
can approach $\approx 90\%$ with an efficiency of about $70\%$
\citep{brunzendorf02}.  In our completeness survey area of
$278\,{\rm deg}^{2}$, a total of approximately 120 unresolved
quasars with $i<19.1$ would not be selected by the SDSS algorithm.
For the entire anticipated SDSS sky area of $\approx 10,000\,{\rm
deg}^{2}$, about $4400$ unresolved $i<19.1$ quasars in the imaging
data will not be selected, compared to about $83,000$ which will
eventually have spectroscopic confirmations. (Resolved quasars and
those fainter than $i=19.1$ will add roughly $15,000$ more objects
to the final SDSS quasar sample.)

The measured completeness value includes the combined color
and radio selection algorithms used by the SDSS.  Of the 2191
spectroscopically confirmed quasars in the main sample, 2165
($98.8\%$) were color selected, and 158 ($7.2\%$) were selected
as optical matches to FIRST radio sources.  Of the radio selected
quasars, 26 ($16.6\%$) were not color selected.  Thus, while only
about $84\%$ of the radio selected quasars were also color selected,
the inclusion of radio selection contributes only about $1\%$ to
the completeness of the main SDSS quasar sample.  These numbers
are consistent with what \citet{ivezic02} found in their analysis
of the radio properties of extragalactic optical SDSS sources.
The completeness for the color and radio selection algorithms
separately are $93.8^{+2.6}_{-3.8}\%$ and $6.8^{+0.9}_{-0.9}\%$
respectively, at the $90\%$ confidence level.

\subsection{Incompleteness Due to Image Defects\label{defects}}
Before quasar target selection is applied to the SDSS photometric
object data, objects with possible image defects or other image
quality problems which may result in unreliable photometry are
rejected \citep{richards02}.  Some of the rejected objects are
quasars which would otherwise have been targeted.  This results
in a quasar survey incompleteness which is independent of the
incompleteness of the color selection algorithm.

The list of image quality flags generated by the $\tt photo$
pipeline is described by \citet{stoughton02}; the combination
of flags used to accept or reject objects before applying the
quasar selection algorithm is described by \citet{richards02}.
There are two stages of image quality tests.  In the first stage,
``duplicate'' objects (such as those that are combinations of several
individual objects blended together, which are also successfully
deblended and recorded separately) and objects with ``fatal'' errors
are rejected.  The most important fatal error is image saturation,
which causes the photometric measurements to be highly unreliable.
Image saturation mainly affects objects significantly brighter
than the $i>15$ limit used by the quasar survey.  The density on
the sky of quasars brighter than this limit is extremely small.
For both of these reasons, image saturation does not significantly
reduce the completeness of the main SDSS survey.

In the second stage, a set of ``non-fatal'' image quality tests is
applied to the objects remaining after the first stage of tests.
Most non-fatal error tests apply to ``children'' --- objects which
have been deblended from a ``parent'' containing at least two
individual objects blended together --- and include checks for very
large magnitude errors, and difficulties in measuring the central
position of an object.  Checks are also made for objects lying
close to known bad CCD columns.  A full description of the tests
is given by \citet{richards02}.  If an object is flagged as having
a non-fatal error, it is rejected for consideration as a quasar
candidate unless it is an optical match to a FIRST radio source.

\begin{figure}
  \plotone{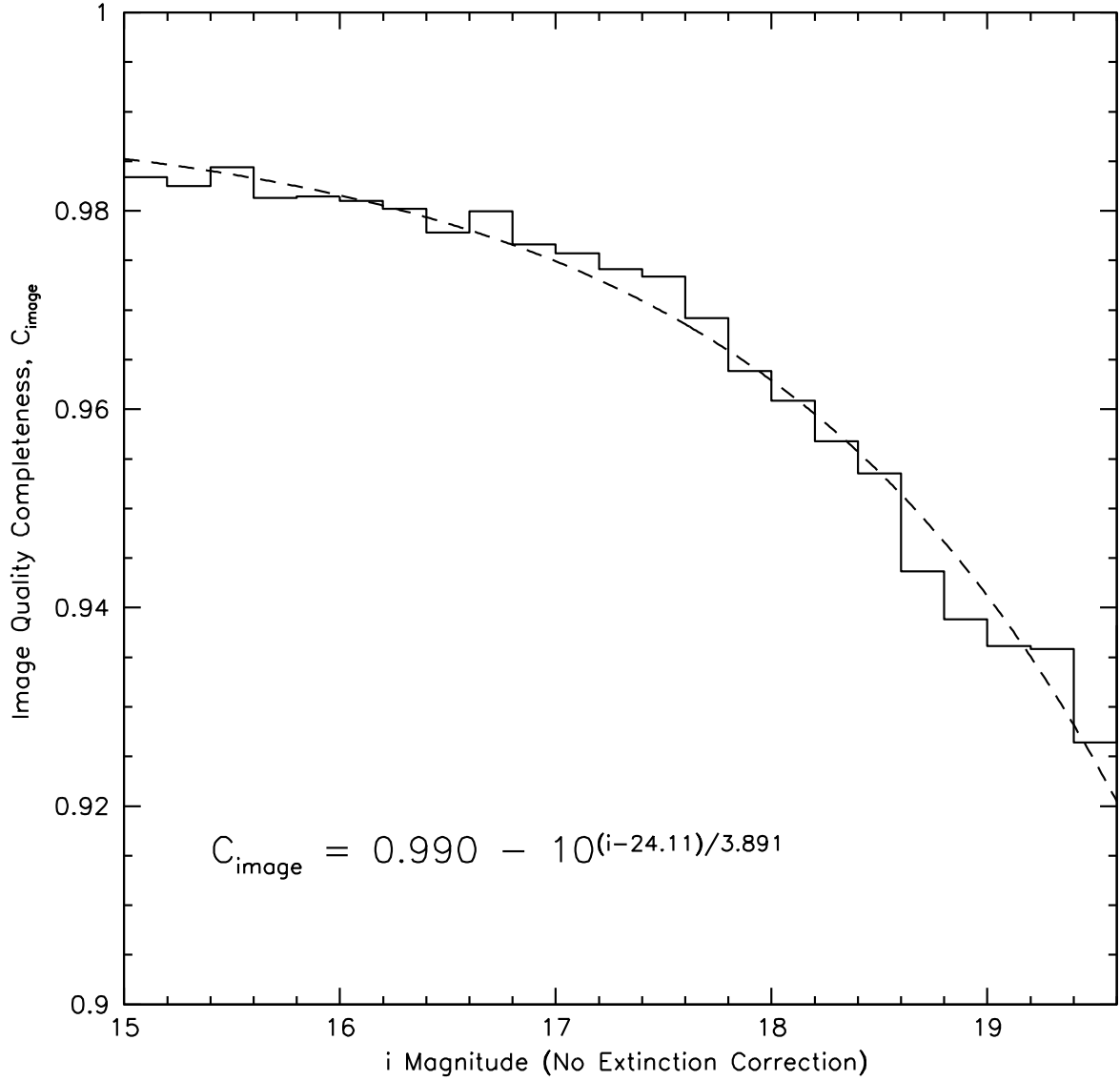}
  \caption{The image quality completeness as a function of $i$ band
  magnitude (solid).  A fit to the data, given by eq.\,(\ref{defectEq}),
  is also shown (dashed).
  \label{defectRatePlot}}
\end{figure}

To measure the incompleteness due to non-fatal image defects,
the image quality tests were applied to unresolved objects that
pass the first stage of (fatal) tests, and have magnitudes within
the $i$ band quasar survey limits.  The fraction of objects which
pass the image quality tests as a function of $i$ band magnitude
(uncorrected for Galactic extinction), i.e.\ the image quality
completeness $C_{image}$, is shown in Fig.\,\ref{defectRatePlot}.
The image quality completeness is above $98\%$ at bright magnitudes,
and drops to about $93\%$ near the faint limit of the quasar survey.
The dependence on magnitude is due to the increased difficulty of
reliably deblending fainter objects.  About $0.5\%$ of the objects
are rejected due to their proximity to bad CCD columns.  The overall
image quality completeness is $96.17\%$ ($628,835$ objects selected
out of $653,859$).  This value is a lower limit,
since some of the objects flagged as having non-fatal errors will
be observed as matches to FIRST radio sources.

Assuming the probability of rejecting an object due to image defects
as a function of flux is described by a power-law over our flux range,
the magnitude dependence of the image quality completeness has the form
\begin{eqnarray}
  C_{image} = 0.990 - 10^{-(i-24.11)/3.89},\label{defectEq}
\end{eqnarray}
where the best fit parameter values are given in the equation.
The fit is shown in Fig.\,\ref{defectRatePlot}, and the equation
is used in \S\,\ref{surfaceDens} for correcting the measurement of
the surface density of quasars.

We have further tested the image defect completeness of the
SDSS quasar survey by matching 2QZ quasars to SDSS imaging data.
There were 7760 matches of 2QZ quasars \citep{croom04} to SDSS
photometry (within $3\arcsec$).  Of those 7760, 1689 are within
the SDSS quasar survey low-$z$ $i$ band magnitude limits and have
UVX quasar colors ($u-g<0.6$ and $g-i>-0.3$).  Of those 1689, 52
($3.1\%$) failed the fatal/non-fatal error tests and were not given
the chance to be selected as SDSS quasar targets --- comparable to
the fraction given above.  Blended objects accounted for 27 of the
52 rejected quasars.

\subsection{Incompleteness in the Spectroscopic Data}
The spectroscopic identification of quasar candidates is the
final stage of the quasar survey, and the last opportunity for
incompleteness to affect the results.  Broken fibers, spectra
with unprocessable regions (bad CCD columns or contamination from
neighboring fibers), and objects with very low S/N
levels introduce a negligibly small incompleteness into the
quasar survey.

It is reported by \citet{abazajian04} that 437 out of 367,360
spectra, or $0.12\%$, in the SDSS DR2 sample were obtained
with fibers that were broken at the time of the observations.
This fraction is negligible compared to the other sources of
incompleteness described in \S\S\ref{algorithm} and \ref{defects}.

There are very few unidentified quasar candidates with spectra,
especially among those brighter than the dereddened limiting $i$
band magnitude of $19.1$.  In a preliminary manual inspection of
all of the quasar candidate spectra in the SDSS Data Release 3
\citep[DR3;][]{abazajian05}, there were less than 20 objects meeting
the magnitude limit that were not identifiable.  There were almost
$29,700$ spectroscopically verified quasars among the same candidate
quasar spectra, which means the incompleteness due to unidentified
spectra is less than $0.07\%$.

We conclude that the incompleteness introduced by spectroscopic
problems or unidentified spectra is at most $0.2\%$.  This is much
smaller than the completeness uncertainty for the quasar selection
algorithm (\S\ref{algorithm}) and the incompleteness due to image
defects (\S\ref{defects}).

\subsection{Quasars with Extended Image Profiles\label{extended}}
The completeness survey addresses only unresolved sources because
a similar survey of extended sources would either be prohibitively
large, or would likely yield few if any untargeted quasars.  However,
a reasonable estimate of the contribution of extended sources to
the full quasar sample can be obtained from the published SDSS
quasar samples, along with the SDSS galaxy spectroscopic samples.

We have analyzed the quasar and galaxy samples in the SDSS Data
Release 3.  Of the quasars selected by the SDSS algorithm that meet
the $i\le19.1$ magnitude limit, $5.7\%$ are classified as extended
by the {\tt photo} imaging pipeline.  Since there is not a complete
spectroscopic sample of extended objects to 19.1 (for the reasons
given above), we have used the SDSS galaxy sample to estimate
the fraction of extended quasars missed by the quasar algorithm.
The galaxy sample is selected according to the algorithm described
by \citet{strauss02}, and nearly every extended object to a limiting
Petrosian magnitude of $r=17.77$ is targeted for spectroscopy.

\begin{figure}
  \plotone{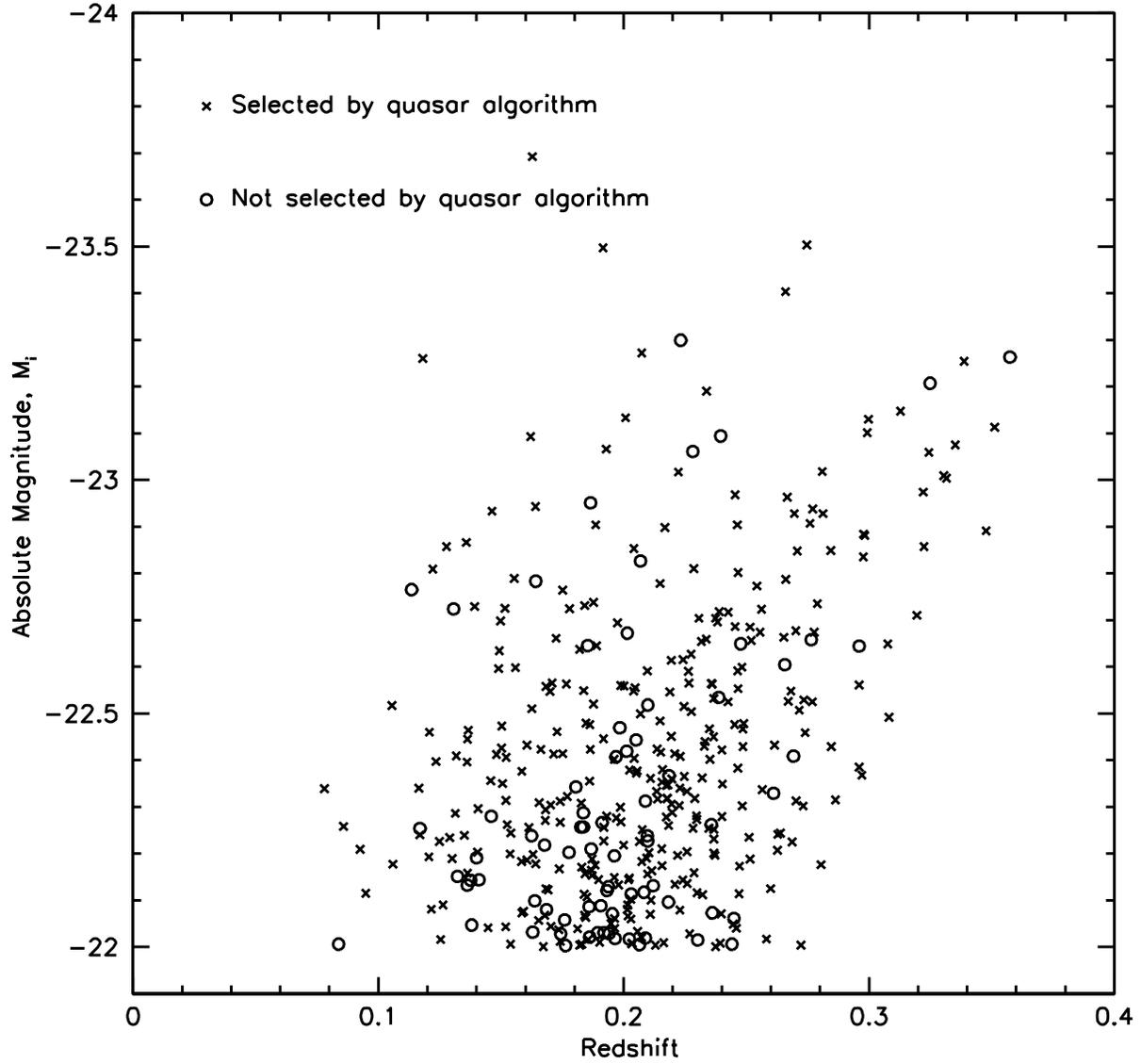}
  \caption{Absolute $i$ band magnitude vs. redshift for quasars
  selected as SDSS galaxy targets.  Those also selected by the quasar
  algorithm are shown with $\times$'s, while those missed by the quasar
  algorithm are shown with open circles.
  \label{zedMiGal}}
\end{figure}

Figure \ref{zedMiGal} shows $i$ band absolute magnitudes and
redshifts of SDSS DR3 quasars that were selected as galaxy
spectroscopic targets.  The vast majority of the quasars have
redshifts less than $0.4$.  Of the quasars with $z>0.4$, most have
images blended with foreground galaxies or stars, and several have
been identified as gravitationally lensed quasars \citep[e.g.][]{inada03}.
In the DR3 main galaxy sample, there are 416 spectroscopically
confirmed quasars (objects with both broad emission lines
and absolute magnitude $M_{i}\le-22.0$).  Of those, 336 were
also selected by the quasar algorithm, for a completeness of
$80.8^{+2.9}_{-3.4}\%$ at the $90\%$ confidence level.  We have
adopted this value for the extended-image quasar selection
completeness, as listed in Table\,\ref{tab2}.  The precise value
depends upon the details of both the quasar and galaxy selection
functions, the SDSS star-galaxy separation algorithm, and the
properties of the quasars and their host galaxies.  However, there
is no evidence that the extended-source quasar completeness depends
upon redshift, luminosity, or apparent magnitude, so the adopted
value is reasonable to use for the current analysis.

\begin{figure}
  \plotone{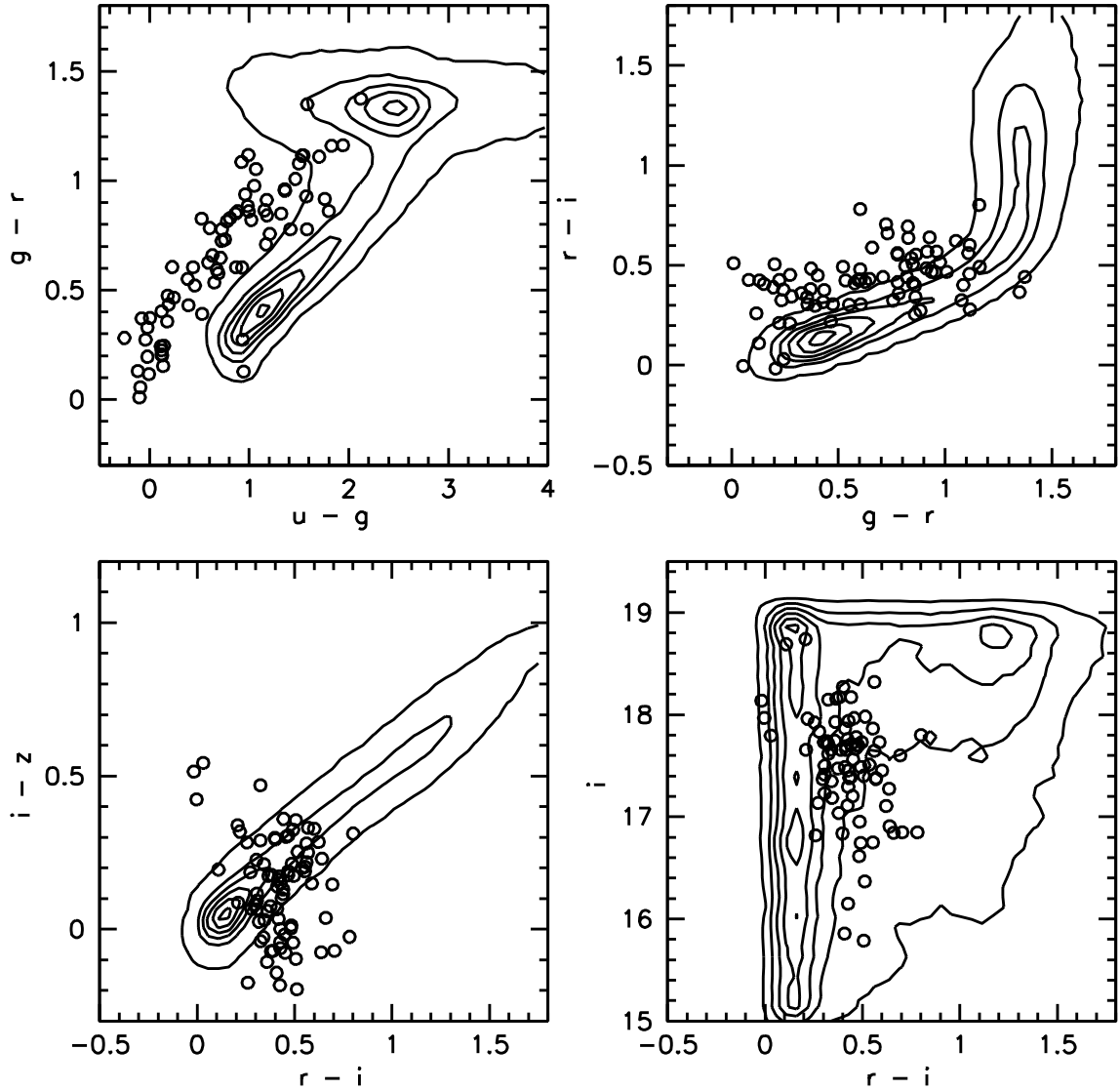}
  \caption{Color-color and color-magnitude diagrams for quasars selected
  as SDSS galaxy targets, but not selected as quasar targets.  The
  contours show the location of the stellar locus as determined from
  the stars in the point source completeness survey.
  \label{cccmGal}}
\end{figure}

Color-color and color-magnitude diagrams of the 80
galaxy-selected quasars missed by the quasar algorithm are shown
in Fig.\,\ref{cccmGal}.  Many of these are missed due to a set of
color cuts imposed on extended objects in order to reject galaxies
less likely to harbor active nuclei.  These cuts have the effect
of removing most of the objects ``above'' and to the ``left''
of the stellar locus in the $ugr$ and $gri$ color-color planes
\citep{richards02}.  The location of the missed extended-image
quasars in color space is due largely to flux contributions from the
host galaxies.  None of the quasars were rejected for having colors
inside any of the exclusion boxes described by \citet{richards02}.

The area spectroscopically covered by the SDSS DR3 is $4188\,{\rm
deg}^{2}$ \citep{abazajian05}.  The density of all quasars with
extended images is then estimated by dividing the number of such
quasars in the DR3 sample selected by the quasar algorithm (1693),
by the extended-image quasar selection completeness, divided by
the DR3 survey area.  The result is, $0.50^{+0.04}_{-0.03},{\rm deg}^{-2}$
at the $90\%$ confidence level.  The density of extended-source
quasars missed by the algorithm is $0.10^{+0.03}_{-0.02}$.  These numbers
are used along with the values for the unresolved quasars in
\S\,\ref{surfaceDens}, to determine the total density of quasars
and the SDSS quasar selection completeness for all quasars.

\subsection{Completeness of the SDSS Quasar Survey and the Surface Density
of Quasars \label{surfaceDens}}
The final completeness of the SDSS quasar survey --- the fraction of
all quasars inside the survey area within the limiting magnitudes
which are finally identified by the SDSS --- is the product of the
completenesses of the image sample, the quasar candidate sample,
and the spectroscopic sample.  Table\,\ref{tab2} summarizes the
measured completenesses at each stage of the SDSS quasar survey.
Multiplying the completeness values, and taking into account
the relative fractions of unresolved and extended-image quasars,
gives a final end-to-end completeness of approximately $89.3\%$ to
within a few percent.

The final completeness value is probably a lower limit, since some
of the missed quasars are actually recovered by other SDSS target
selection algorithms, such as matching to the ROSAT All-Sky Survey
\citep[RASS,][]{voges99,voges00} sources \citep[e.g.][]{anderson03},
and various types of star candidate selection \citep{stoughton02}.
For example, there are 16 quasars in the completeness survey
image area which were not targeted by the main quasar algorithm,
but which were targeted by another SDSS algorithm.  Objects such as
these are no more likely to have been targeted by the sparse-sampled
completeness survey than any other random source (and none of the
16 were targeted).  These objects are ignored for all statistical
purposes; their numbers cannot be used to reliably measure an
improved completeness, because they are not assigned spectroscopic
fibers in the SDSS in a predictable way.  For further discussion
of the contribution of these objects to the quasar sample, see
\S\,\ref{discussion}.

The surface density of all quasars to the limiting magnitude
can be estimated by weighting each of the detected quasars by
the completenesses at each stage of the survey, and accounting
for the contribution of extended image quasars, and the number of
quasars expected from the unverified candidates.  Because some of
the completenesses are are functions of magnitude, each quasar has
to be weighted according to its magnitude.  The estimated number
of quasars to $i=19.1$ in the $277.6\,{\rm deg}^{2}$ area of the
completeness survey is 2828, for a surface density of
$10.19\,{\rm deg}^{-2}$.  

\begin{figure}
  \plotone{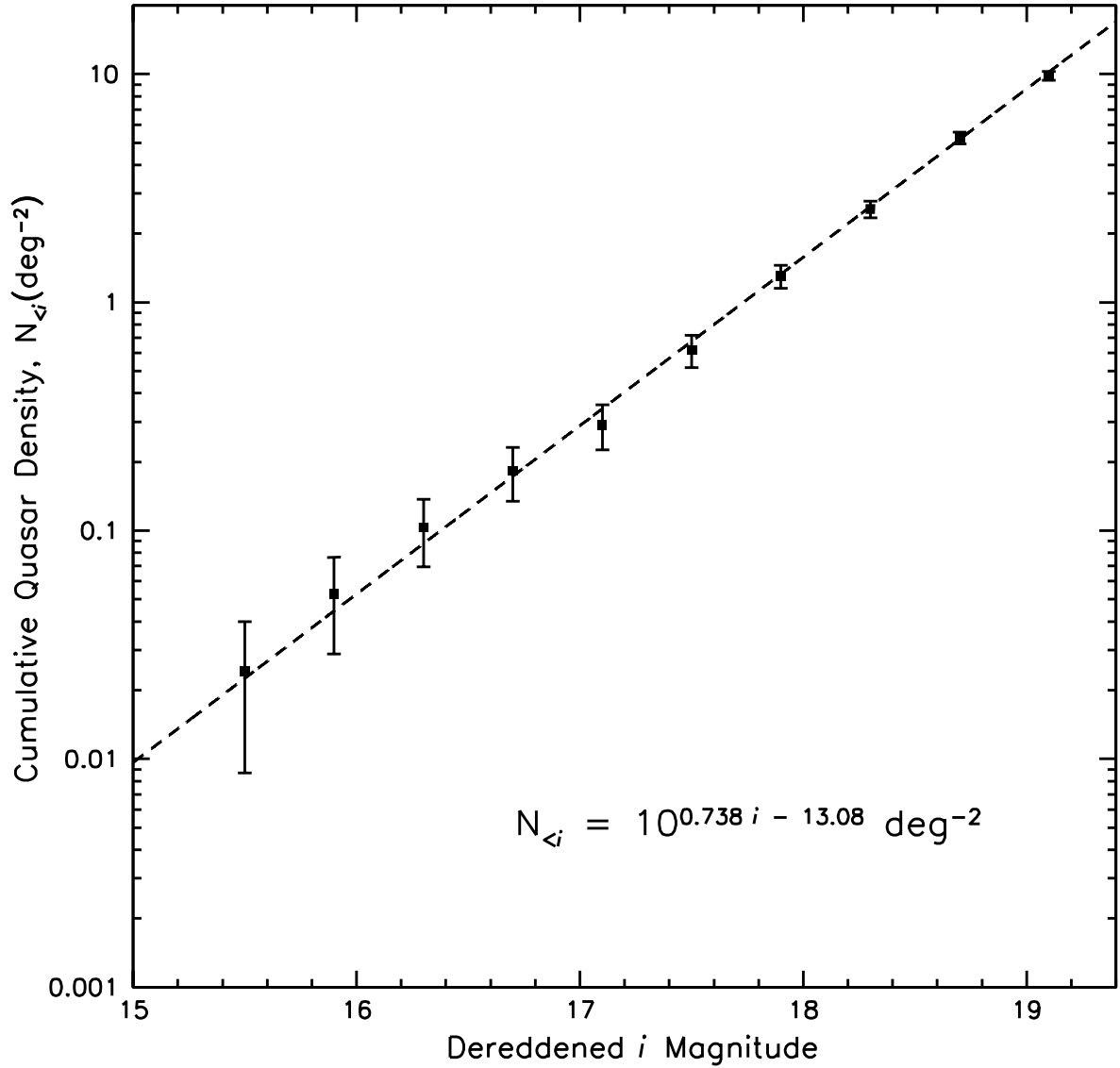}
  \caption{Cumulative surface density of quasars as a function of 
  dereddened $i$ band magnitude, corrected for the SDSS incompleteness.
  A single exponential fit (dashed line) is also shown.  The parameter 
  values were determined using a maximum likelihood fit to the
  individual detected quasars; the binned points are for display
  purposes only.
  \label{magDens}}
\end{figure}

The cumulative surface density of quasars as a function of dereddened
$i$ magnitude is shown in Fig.\,\ref{magDens}.  The number of quasars
up to the limiting magnitude in each bin was estimated using the same
weighting as for the full sample.  The error bars were found by 
taking the square root of the number of detected quasars in each bin
and scaling by the average weights.  The figure also shows a maximum
likelihood fit to the number density assuming an exponential form
\begin{eqnarray}
  N_{<i} = N_{0}10^{\alpha i},
\end{eqnarray}
where the maximum likelihood values of the parameters are
$\log N_{0} = -13.08\pm0.73$ and $\alpha = 0.738\pm0.038$ at the $90\%$
confidence level.  The likelihood distribution of the $i$ magnitudes
of the individual detected quasars was used to determine the best
fit; the binned points are used for display purposes only.

The exponential form fits the number density well across the
observed magnitude range.  There is no evidence for a ``break''
or strong curvature in the data, as has been reported in some
studies \citep[e.g.][]{croom01}.  However, the limiting magnitude
of the completeness survey is probably simply too bright to detect
any deviations from a pure exponential.  In any case, it is not the
goal of the current study to determine the precise form of the quasar
surface density function, since we do not know with precision how
the final completeness varies with magnitude.  However, the final
completeness and the surface density up to the limiting magnitude
{\em are} now accurately measured quantities.

\section{Discussion\label{discussion}}
The overall completeness of the SDSS quasar selection algorithm
and the survey itself are quite high, and surpass the survey goals.
We would still like to understand how the completeness varies as a
function of redshift or other parameters.  Although the sample of
quasars not found by the quasar algorithm is small (10), there are
a number of general statements that may be made about why these
were not targeted.  A more detailed discussion of each quasar is
given in \S\,\ref{notes}.

All of the missed quasars have colors consistent with the stellar
locus in the $ugri$ (low-$z$) color cube.  Six of the ten quasars
are significantly redder than average for their redshifts, which
is the main reason their colors lie closer to the stellar locus.
Of the six red quasars, BAL systems were detected in three of them.
It is known that BALQSOs tend to be redder than non-BALQSOs
\citep[e.g.][]{reichard03,brotherton01}.  The number of BALQSOs
in the sample is too small to determine if their density among
missed quasars is significantly different than for all quasars;
however it does not appear that they dominate the missed quasars.

Red or reddened quasars appear to comprise a large fraction of
the quasars missed by the SDSS algorithm.  This does not mean,
however, that a large fraction of apparently red quasars is missed
by the algorithm.  In fact, the SDSS selects quasars with a wide
range of colors relative to the mean colors at every redshift
\citep{richards01,richards03}.  Our study is concerned with the number of
quasars with {\em apparent} magnitudes brighter than $i=19.1$, but
it does not address the fraction of quasars which may be fainter
than the survey limit due to extinction, either intrinsic to the
quasars or along the sightlines.  The issue of quasar reddening
in the SDSS sample is explored in detail by \citet{richards03}
and \citet{hopkins04}.  So-called type\,II quasars, which exhibit
narrow high-ionization emission lines due to nuclear obscuration,
found in the SDSS are discussed by \citet{zakamska03}.

The missed quasar sample size is too small to determine whether
its redshift distribution is significantly different from the main
SDSS sample.  The close grouping of 4 quasars near $z=2.9$ is not
surprising since $z=2.9$ is close to where the average quasar color
crosses the blue end of the stellar locus.  Objects with colors
common to both blue stars and average quasars are sparsely sampled
(from the mid-$z$ inclusion region) at a fixed rate in the SDSS
selection, and simple corrections can be applied to quasar counts
with those colors.

Quasars in the full SDSS dataset may also be selected as matches
to ROSAT X-ray sources \citep{anderson03}, or as various classes of
stellar candidates \citep{stoughton02}.  (Quasars are also selected
by the SERENDIPITY algorithms, but $99.7\%$ of those quasars are
also selected by other algorithms, so we omit a discussion of the
SERENDIPITY selected quasars.)  Of the 2191 main sample quasars, 87
($4.0\%$) were also X-ray selected, and 19 ($0.9\%$) were selected
by algorithms designed to search for interesting types of stars.
In the completeness survey area, there are an additional 7 X-ray
selected and 6 stellar selected quasars which were neither color nor
radio selected.  The color and radio algorithms select almost $93\%$
(87 of 94) of the X-ray selected quasars. A smaller fraction, $68\%$
(17 of 25), of the stellar selected quasars were selected in the
main sample.  The 13 X-ray or stellar selected objects missed in
the main quasar sample, if included, would add less than $1\%$
to the completeness of the survey.

Statistical measures of the distributions of quasars, such as the
luminosity function and correlation function, which rely on the SDSS
data must take the survey completeness into account.  The results of
this program show that the overall completeness of the SDSS quasar
survey is quite high.  However, the results at this point cannot
be used to determine in detail how the statistical measures should
be corrected as functions of various parameters, such as redshift
or luminosity.  This is due in part to the high completeness of
the survey itself -- even out of some $20,000$ objects, so few
missed quasars were detected that the parameter dependencies cannot
be determined.  Those dependencies will be addressed in a future
paper (Richards et~al., in preparation) using simulated spectra.
For now, the results at least qualitatively suggest that quasars
missed by the SDSS have a redder color distribution, and that they
may disproportionately occupy certain narrow redshift ranges.

\section{Conclusions\label{conclusions}}
In this paper we have empirically quantified the global completeness
of the SDSS quasar survey to a limiting dereddened magnitude of
$i=19.1$.  Of almost $20,000$ spectroscopic identifications of
randomly selected objects, only 10 quasars were found which were
not targeted by the main (color and radio) SDSS quasar selection
algorithm.  The missed quasars either had redder than average colors
which made them consistent with the stellar locus, or they fell into
a narrow redshift range near $z=2.9$ which is sparsely sampled by
the SDSS algorithm.  The completeness of the selection algorithm
is approximately $95\%$.   Accounting for objects rejected due to
image defects, unidentifiable spectra, and extended sources, the
overall completeness of the SDSS quasar survey is approximately
$89\%$.  After accounting for the completeness, the total density
of unresolved quasars on the sky to $i=19.1$ is estimated to be
$10.2\,{\rm deg}^{-2}$.  While it is not yet determined how the
completeness varies with redshift, luminosity, or other quasar
parameters, the precise determination of the global completeness
will allow for more accurate determinations of the quasar luminosity
function, and other statistical measures of quasar distributions.

\acknowledgments
We thank the anonymous referee for a number of helpful suggestions
and tests.  This work was supported in part by National Science
Foundation grants AST 03-07582 (D.P.S.), AST 00-71091 and AST
03-07409 (M.A.S.).

Funding for the creation and distribution of the SDSS Archive has
been provided by the Alfred P. Sloan Foundation, the Participating
Institutions, the National Aeronautics and Space Administration,
the National Science Foundation, the U.S. Department of Energy,
the Japanese Monbukagakusho, and the Max Planck Society. The SDSS
Web site is http://www.sdss.org/.

The SDSS is managed by the Astrophysical Research Consortium (ARC)
for the Participating Institutions. The Participating Institutions
are The University of Chicago, Fermilab, the Institute for Advanced
Study, the Japan Participation Group, The Johns Hopkins University,
the Korean Scientist Group, Los Alamos National Laboratory, the
Max-Planck-Institute for Astronomy (MPIA), the Max-Planck-Institute
for Astrophysics (MPA), New Mexico State University, University
of Pittsburgh, Princeton University, the United States Naval
Observatory, and the University of Washington.


\onecolumn
\clearpage
%
%
%
%
%
%
%
%
%
%

\clearpage
%
\begin{deluxetable}{rrrrrrrrrrc}
\tablecolumns{11}
\tabletypesize{\scriptsize}
\rotate
\tablecaption{Spectroscopically identified quasars. \label{tab1}}
\tablewidth{0pt}
\tablehead{
  \colhead{No.} &
  \colhead{SDSS J} &
  \colhead{$\alpha_{J2000}$} &
  \colhead{$\delta_{J2000}$} &
  \colhead{Redshift} &
  \colhead{$i$\tablenotemark{a}} &
  \colhead{$u-g$\tablenotemark{a}} &
  \colhead{$g-r$\tablenotemark{a}} &
  \colhead{$r-i$\tablenotemark{a}} &
  \colhead{$i-z$\tablenotemark{a}} &
  \colhead{Selection} \\
  \colhead{} &
  \colhead{} &
  \colhead{(deg)} &
  \colhead{(deg)} &
  \colhead{} &
  \colhead{(mag)} &
  \colhead{(mag)} &
  \colhead{(mag)} &
  \colhead{(mag)} &
  \colhead{(mag)} &
  \colhead{}
}
\startdata
 1 & 003517.95$+004333.7$\tablenotemark{b} &    8.82481 &    0.72604 & 2.898
   &  19.03 &   1.43 &   0.44 &   0.33 &   0.04 & All Color\\
 2 & 003719.85$+011114.6$ &    9.33269 &    1.18740 & 0.401
   &  18.31 &   0.88 &   0.45 &   0.24 &   0.38 & Locus Out\\
 3 & 013011.42$+001314.6$ &   22.54757 &    0.22072 & 1.054
   &  18.90 &   1.35 &   0.82 &   0.31 &   0.14 & All Color\\
 4 & 031732.20$+000209.7$\tablenotemark{b} &   49.38418 &    0.03603 & 2.324
   &  18.72 &   1.32 &   1.52 &   0.69 &   0.31 & Locus Out\\
 5 & 032228.99$-005628.6$ &   50.62078 &   -0.94128 & 2.974
   &  19.07 &   1.05 &   0.45 &   0.19 &   0.00 & Locus Out\\
 6 & 034629.02$+002337.7$ &   56.62091 &    0.39380 & 2.770
   &  19.03 &   1.00 &   0.16 &  -0.01 &   0.19 & All Color\\
 7 & 204626.11$+002337.7$ &  311.60878 &    0.39381 & 0.332
   &  17.63 &   1.08 &   0.76 &   0.05 &   0.69 & All Color \\
 8 & 215241.89$-001308.7$\tablenotemark{b} &  328.17453 &   -0.21908 & 1.527
   &  18.12 &   1.20 &   0.41 &   0.12 &  -0.08 & Locus Out\\
 9 & 221936.37$+002434.1$ &  334.90155 &    0.40947 & 2.854
   &  18.59 &   0.88 &   0.12 &   0.02 &   0.03 & All Color \\
10 & 231937.64$-010836.1$ &  349.90683 &   -1.14335 & 0.770
   &  18.91 &   0.70 &   0.30 &   0.10 &   0.12 & All Color\\
\enddata
\tablenotetext{a}{Corrected for Galactic reddening according to
  \citet{schlegel98}.  Typical uncertainties are less than $0.03$
  magnitudes.}
\tablenotetext{b}{Broad absorption line quasar.}
\end{deluxetable}

%
\begin{deluxetable}{rr}
\tablecolumns{11}
\tabletypesize{\scriptsize}
\tablecaption{SDSS quasar survey completeness summary. \label{tab2}}
\tablewidth{0pt}
\tablehead{
  \colhead{Selection} &
  \colhead{Completeness\tablenotemark{a}}
}
\startdata
 Image selection / clean images & $96.17^{+0.04}_{-0.04}$\% \\
 Color selection only & $93.8^{+2.6}_{-3.7}$\% \\
 Radio selection only & $6.8^{+0.2}_{-0.3}$\% \\
 Color and radio selection & $94.9^{+2.6}_{-3.8}$\% \\
 Extended sources\tablenotemark{b} & $80.8^{+2.9}_{-3.4}$\% \\
 Identifiable spectra & $>99.8\%$ \\
 Total completeness to $i=19.1$ & $89.3\%$ \\
\enddata
\tablenotetext{a}{Fraction of true quasars which would pass selection
criterion.}
\tablenotetext{b}{Fraction of true quasars with extended image profiles
that would be selected by the quasar algorithm.}
\end{deluxetable}

\end{document}